\documentclass[a4paper,fleqn,usenatbib,useAMS]{mnras}

\usepackage{newtxtext,newtxmath}

\usepackage[T1]{fontenc}
\usepackage{ae,aecompl}

\usepackage{graphicx}	
\usepackage{amsmath}	
\usepackage{amssymb}	

\newcommand{\rcusp}{$R_{\rm cusp}$}
\newcommand{\rfold}{$R_{\rm fold}$}

\title[Flux-ratio anomalies from discs]{
Flux-ratio anomalies from discs and other baryonic structures in the Illustris simulation}
\author[Hsueh et al.]{Jen-Wei Hsueh$^{1}$\thanks{E-mail:
jwhsueh@ucdavis.edu}, 
Giulia Despali$^{2}$, 
Simona Vegetti$^{2}$, 
Dandan Xu$^{3}$, 
\newauthor Christopher D. Fassnacht$^{1}$,
 and R. Benton Metcalf$^{4,5}$\\
$^{1}$Physics Dept., University of California, Davis, 1 Shields Ave. Davis, CA 95616, USA\\
$^{2}$Max Planck Institute for Astrophysics, Karl-Schwarzschild-Strasse 1, D-85740 Garching, Germany\\
$^{3}$ Heidelberg Institute for Theoretical Studies, Schloss-Wolfsbrunnenweg 35, D-69118 Heidelberg, Germany\\
$^{4}$ Dipartimento di Fisica e Astronomia, Universita di Bologna, Via Gobetti 93/2, 40129 Bologna, Italy\\
$^{5}$ INAF-Osservatorio Astronomico di Bologna, Via Gobetti 93/3, 40129 Bologna, Italy \\
}

\begin{document}


\pagerange{\pageref{firstpage}--\pageref{lastpage}} \pubyear{2017}

\maketitle

\label{firstpage}

\begin{abstract}
The flux ratios in the multiple images of gravitationally lensed quasars can provide evidence for dark matter substructure in the halo of the lensing galaxy if the flux ratios differ from those predicted by a smooth model of the lensing galaxy mass distribution.  However, it is also possible that baryonic structures in the lensing galaxy, such as edge-on discs, can produce flux-ratio anomalies. In this work, we present the first statistical analysis of flux-ratio anomalies due to baryons from a numerical simulation perspective. We select galaxies with various morphological types in the Illustris simulation and ray-trace through the simulated halos, which include baryons in the main lensing galaxies but exclude any substructures, in order to explore the pure baryonic effects. Our ray-tracing results show that the baryonic components can be a major contribution to the flux-ratio anomalies in lensed quasars and that edge-on disc lenses induce the strongest anomalies.
We find that the baryonic components increase the probability of finding high flux-ratio anomalies in the early-type lenses by about $8\%$ and by about $10 - 20 \%$ in the disc lenses. The baryonic effects also induce astrometric anomalies in $13\%$ of the mock lenses.
Our results indicate that the morphology of the lens galaxy becomes important in the analysis of flux-ratio anomalies when considering the effect of baryons, and that the presence of baryons may also partially explain the discrepancy between the observed (high) anomaly frequency and what is expected due to the presence of subhalos as predicted by the CDM simulations.

\end{abstract}


\section{Introduction}

One of the key probes for investigating the nature of dark matter is a well-constrained determination of the mass function of substructure associated with galaxy-scale halos.  Strong gravitational lensing is an especially powerful tool for detecting substructure in distant galaxies, which is achieved via two main approaches, namely the gravitational imaging technique and the analysis of flux-ratio anomalies. The gravitational imaging technique \citep{K05,V09} focuses on systems in which the background galaxy is lensed into a long arc or Einstein ring.  Substructures that are located close to the arcs or ring produce small astrometric perturbations to the lensed emission, which can be detected as a residual when comparing the observed surface brightness distribution to that predicted by a smooth (i.e., without substructure) mass lens model. Several substructures with $10^8 - 10^9 M_{\sun}$ have now been detected with this technique \citep{V10,V12,Hezaveh16}. 
With the gravitational imaging technique, the projected position and mass of the substructure can be determined. The minimum detectable substructure mass is determined by the source size, surface brightness structure of the source, and the angular resolution of the imaging.  Thus, this technique requires high-resolution imaging.  The inferred substructure abundances from current samples that have been analyzed with the gravitational imaging technique are marginally consistent with those predicted by dark-matter-only and hydrodynamical simulations although the sample size is limited \citep{V14a,Despali2016}.

The other approach, the analysis of flux-ratio anomalies, was first proposed by \citet{Mao1998} and \citet{metcalf01}.  In this case, the targeted systems show multiple lensed images of a background active galactic nucleus (AGN).  Deviations between the ratios of the observed fluxes of the lensed AGN and those predicted by smooth lens  models -- called ``flux-ratio anomalies'' -- can be caused by small-scale structure in the halo of the lensing galaxy such as dark matter substructure.  In particular, in four-image lens systems the flux ratios of ``merging images'' (i.e., those that are very close together on the sky) follow relations that are nearly universal for smooth mass models. These flux ratios are sensitive to perturbations in the lensing potential. To reliably quantify the perturbation effect from substructures, it is necessary to obtain flux measurements that are as free as possible from non-gravitational effects.  In practice, this often means fluxes that have been measured at radio wavelengths and with long-term monitoring (in order to average out the intrinsic variation of the lensed AGN). This is because the radio emission region in quasars is much larger than the Einstein radius from stars in the foreground lensing galaxy, and thus the radio fluxes are free from the stellar microlensing that often seen at optical wavelengths \citep[but also see][for a rare case.]{koopmans00} Also, observing at radio wavelengths can minimize  dust extinction, which affects flux ratios at shorter wavelengths. Although propagation effects such as interstellar scattering and free-free absorption can influence the radio wavelength flux ratios, their frequency dependence are well understood \citep{Mittal07,winn04}.

\citet{Dalal2002} presented the first statistical results from the flux-ratio technique, using a sample of  seven radio-loud lensed quasars.  Their inferred substructure abundance is consistent within the errors with the predictions from cold dark matter (CDM) cosmology, although the monitoring in the flux measurements was not done yet at that time. In the follow-up work of \citet{KD04}, they explore alternative sources of flux-ratio anomalies and conclude that these alternatives are less likely to be the cause of the observed anomalies. The assumption that the flux-ratio anomalies in radio-loud lenses are generated solely by the substructure then became the standard in subsequent  studies, such as for example, \citet{Bradac02}, \citet{2002ApJ...567L...5M}, \citet{Dobler2006}, \citet{Fadely2012} and \citet{N14} \citep[however see][ for the the case of no detection.]{Nierenberg2017} 
However, it is not entirely clear whether the CDM substructure scenario is fully compatible with current flux anomaly observations. Numerical studies based on CDM simulations (e.g., \citealt{Mao04,Maccio06,Chen11,2012MNRAS.419.3414M,Xu09,Xu15}) show that the predicted CDM substructure population is not sufficient to reproduce the high flux-ratio anomaly strengths currently observed. This result suggests that assuming substructures are the only source of flux-ratio anomalies may be over-simplified.

Complex baryonic structures in the lens galaxies may provide a viable explanation for the observed high anomaly strength. For example, \citet{Moller03} and \citet{Quadri2003} have discussed the influence of disc structure on flux ratios and
 \citet{Hsueh2016,Hsueh17} have recently shown that with the inclusion of an edge-on disc in the lens model, motivated by high-resolution imaging, the observed flux ratios and positions of lensed images can be successfully reproduced without the need for substructure.
\citet{Gilman2017} have also shown that in elliptical lenses, $10 - 15 \%$ of anomalies are generated from the baryonic structures in the lens galaxies \citep[see also][from the simulation perspective]{Xu10}. Moreover, line-of-sight structures can also contribute to the flux-ratio anomalies \citep{Metcalf05,Xu12,McCully14}. These alternative sources of flux-ratio anomalies may explain why the CDM substructure abundance predicted by N-body simulations at galactic- and group-scales cannot reproduce the observed high frequency of flux-ratio anomalies \citep{Xu09,Xu15}. 

In this paper, we explore the effect of stellar discs and other baryonic structures on flux-ratio anomalies using a state-of-the-art cosmological hydrodynamical simulation -- the Illustris simulation \citep{Vo2014a,Vo2014b,Genel2014,Nelson2015}. This paper is organized as follows. In section 2, we describe the  simulation, the criteria for selecting lenses we use for this work and the final sample of galaxies selected from the simulation. In section 3, we present the ray-tracing results for smooth mass models and the simulated lenses. In section 4, we discuss the flux-ratio and astrometric anomalies in our mock lenses. Finally, in section 5, we summarize the results in this work.

\section{Simulated lenses}

The goal of this paper is to investigate the effect of edge-on discs and other baryonic structures on the flux ratios of gravitationally lensed quasars. Recent hydrodynamical simulations are able to produce a realistic population of galaxies, with a variety of morphological types, and so constitute the ideal tool for our purposes. The first step of our work is then to construct a sample of simulated lens galaxies from the Illustris simulation. The Illustris Project is a series  of hydrodynamical simulations of cosmological volumes that follow the evolution of dark matter, cosmic gas, stars, and super-massive black holes from a starting redshift of $z  = 127$  to the  present time.  At the highest resolution level, which we used in this work, the simulation covers a cosmological volume of $(106.5~{\rm Mpc})^3$ and has a mass resolution of $6.26\times 10^6 M_{\odot}$ and $1.26\times 10^6 M_{\odot}$ for dark matter and baryons, respectively.  The simulations were run using the recent  moving-mesh   AREPO  code  \citep{springel10} and the adopted
cosmological model has $\Omega_{m}=0.2726$, $\Omega_{\Lambda}=0.7274$, $\Omega_{b}=0.0456$, $h=0.704$ and $\sigma_{8}=0.809$, consistent with the WMAP-9 measurements \citep{wmap9}.
 More details on the simulation and the implementation of baryonic physics can be found in \citet{Vo2014a,Vo2014b,Genel2014} and \citet{Nelson2015}. 

\newpage

In order to select a sample of disc galaxies compatible with the available observational data \citep[e.g.][]{CLASS1,CLASS2,JVAS1,JVAS2,JVAS3,Chiba2005,Minezaki2009}, we focus on objects at redshift $z_l = 0.6$. At first, we apply the following general selection criteria: (i) a galaxy must be the central galaxy of the considered host halo; (ii) the dark-matter halo must be less massive than $5 \times 10^{13}~ M_{\sun}$ since above this halo mass, galaxies are in a cluster environment;
(iii) a galaxy must have a stellar mass $M_* > 10^{10}~ M_{\sun}$: this corresponds to $\sim$10,000 stellar particles  which is sufficient to  reliably estimate the broad galaxy morphology (`elliptical' vs. `disc') by resolving the radial surface brightness and calculating the internal kinematics. In addition, we apply two further selection criteria - one based on morphology and the other on stellar kinematic information - in order to identify a disc galaxy sample and its early-type counterpart. We describe these selection criteria in more detail in the following sections.

\subsection{Morphological criteria} \label{ssec:mor}
\begin{figure*}
\centering
\includegraphics[scale=1.0]{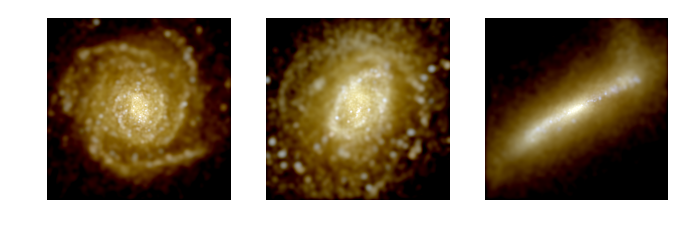}
   \caption{
   Synthesized images of Illustris disc galaxies with different inclination angles ($i$). \textsl{Left}: $i = 7^{\circ}$; \textsl{Center}: $i = 50^{\circ}$; \textsl{Right}: $i = 80^{\circ}$. The images are produced by combining the surface brightness maps in the rest-frame SDSS $g$, $r$ and $i$ filter bandpasses  \citep[see][for more details.]{Xu17}} \label{fig:inc}
\end{figure*}

For the morphology-based selections, the galaxy classifications and S\'{e}rsic profile fitting results were taken from \citet{Xu17}, where the morphological type of each galaxy was determined by fitting both the de Vaucouleurs profile and the exponential profile to the radial surface brightness distribution of the elliptical isophotes. Figure \ref{fig:inc} presents the synthesized images of some examples of disc galaxies in the Illustris simulation. The images are produced by combining the surface brightness maps in the rest-frame SDSS $g$, $r$ and $i$ filter bandpasses \citep[see][for more details.]{Xu17} A galaxy is classified as an early-type if the de Vaucouleurs profile provides a better fit, while if the exponential profile is a better fit then the galaxy is classified as a late-type. Thus, to further restrict our disc galaxy sample, we also require that the selected disc galaxy must also have a S\'{e}rsic index of $n < 2$. Galaxy types defined in this way are referred to as ``morphology-selected'' types.

\begin{figure*}
\centering
\includegraphics[scale=0.50]{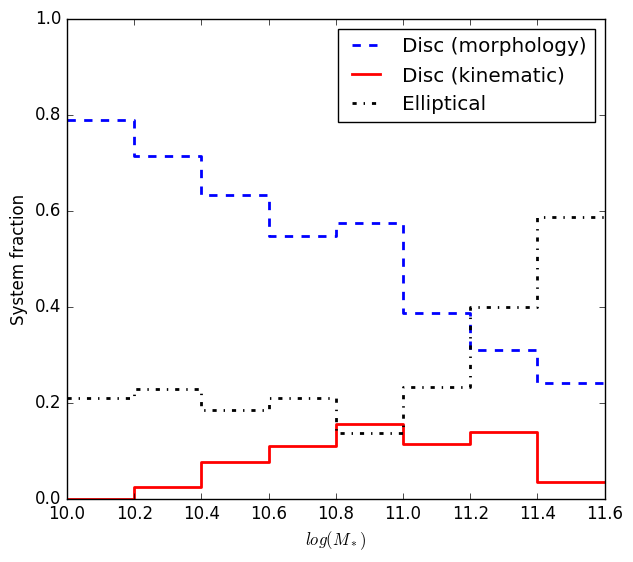}
\includegraphics[scale=0.50]{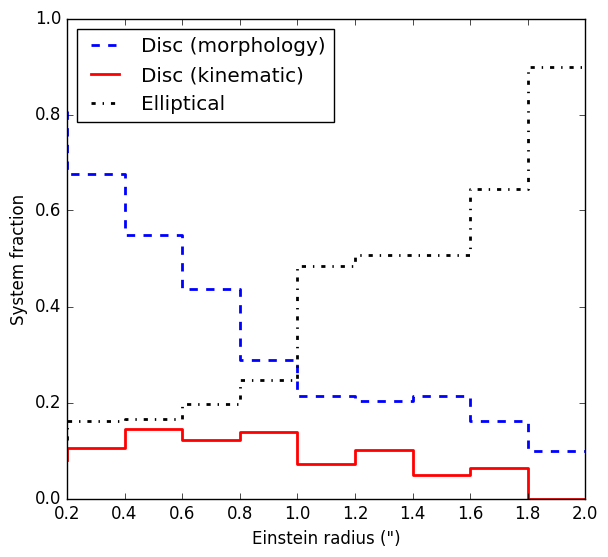}
   \caption{
   \textit{Left}: Lens system fraction versus stellar mass ($M_{\sun}~h^{-1}$) in Illustris at redshift $z = 0.6$. The morphology-selected disc lenses are shown in the blue dashed line, kinematics-selected disc lenses are shown in the red solid line, and elliptical lenses are shown in the black dot-dashed line. \textit{Right}: Lens system fraction versus Einstein radius (in arcsecond) with source redshift assigned as $z_s = 2.0$. Note that some lens galaxies satisfy both the morphological and kinematic criteria, and some galaxies do not meet either disc or elliptical criteria. }\label{fig:sm_hist}
\end{figure*}

\subsection{Kinematic criteria} \label{ssec:kine}
From a kinematic point of view, a stellar disc is a group of stars that co-rotates around the  symmetric axis (we will call this symmetric axis the ``z-axis,'' for simplicity). By looking into the kinematic properties of each stellar particle, a simulated galaxy can be decomposed into a thin disc, a thick disc, and a bulge. 
This dynamical decomposition is usually done in the parameter space of the z-component of the specific angular momentum, $J_z$, and the specific binding energy, $E$, of each star \citep{Abadi03}. On the $J_z - E$ plane, stellar and gaseous components co-rotate, which defines a solid curve $J_{circ}(E)$ \citep[see Fig. 2 in][]{Abadi03} that represents the co-rotating circular orbits in the disc as a function of specific binding energy. The circularity parameter $\epsilon_z$ of each star is defined as the ratio between its  $J_z$ and corresponding $J_{circ}(E)$, showing how closely a stellar or gaseous component follows the circular motion of the disc.  In general, the gaseous disc and the thin disc stellar components have a sharp distribution of $\epsilon_z$ peaked about unity. Based on the $\epsilon_z$ distribution, all stars within the ``luminous'' radius can be decomposed into three different dynamical components: a bulge, a thin disc, and a thick disc.  For the Illustris simulation, the quantity $\epsilon$ has been calculated for all particles and for each galaxy the following quantities are provided  \citep[see][for more details]{Genel2015}:
\begin{itemize}
\item the fraction of stars with $\epsilon_z>0.7$, which can be used to 
  identify the thin disc stars - those with significant (positive)
  rotational support
\item the fraction of stars with $\epsilon_z<0$, multiplied by two, which
  in turn defines the mass fraction of the bulge.
\end{itemize}
Following \citet{Teklu2015}, galaxies with more than 40 percent of stars in the thin disc (above-mentioned fraction 1) are considered as ``kinematics-selected'' disc galaxies. On the other hand, galaxies with more than 60 percent of stars in the bulge (above-mentioned fraction 2) are considered as ``kinematics-selected'' elliptical/early-type galaxies.

\subsection{Disc and elliptical lens samples in this work}

Here we describe the final galaxy samples used in this work and we summarize their distributions.
Figure \ref{fig:sm_hist} shows the dependence on stellar mass (left panel) and on Einstein radius\footnote{ The Einstein radius $R_{\rm E}$ of each galaxy is calculated as the projected radius within which the mean surface mass density $\bar{\Sigma}(\leqslant R_{\rm E})$ is equal to the critical density $\Sigma_{\rm cr}=\frac{C^2}{4\pi G} \frac{D_{\rm s}}{D_{\rm ds} D_{\rm d}}$, where $D_{\rm d}$, $D_{\rm s}$ and $D_{\rm ds}$ are the angular diameter distances from the observer to the lens, to the source, and from the lens to the source, respectively.} (right panel) of the fractions of galaxies that are defined as one of the following types: morphology-selected discs, kinematics-selected discs, and elliptical galaxies, at $z=0.6$ from the Illustris simulation. The source redshift is fixed at $z_s = 2.0$ to maximize the Einstein radius for the simulated lenses. 
Note that the histograms in Figure \ref{fig:sm_hist} do not sum up  to 1.0 because some lens galaxies satisfy both the morphological and kinematic criteria, and some other galaxies do not meet either criteria. However, it is still clear from the distributions that the majority of lens systems with small Einstein radii are produced by disc galaxies, as expected \citep[e.g.][]{Turner84}.
We note that the selection based on kinematics is more stringent than the morphological one, as it requires the formation of well-established discs; as a consequence, the number of kinematics-selected galaxies is small, especially in the low mass regime \citep[also see ][]{Bottrell2017}. In our final disc sample, we only include those galaxies that satisfy both criteria.

For each selected disc galaxy, we calculate an inclination angle, which is defined as the angle between the line-of-sight and the major rotational axis, i.e., the ``z-axis''. From the visualization of simulated galaxies, we further define a subsample of edge-on discs whose inclination angles are larger than 80 degrees and a subsample of face-on discs whose inclination angles are smaller than 50 degrees. Figure \ref{fig:inc} presents the synthesized images of example disc galaxies of different inclination angles from the Illustris simulation. 

Our final elliptical sample includes galaxies that simultaneously satisfy: (1) having a better fit to the de Vaucouleurs profile in surface brightness, (2) S\'{e}rsic index larger than two, and (3) more than 60 percent of stars in the bulge.  A summary of the selection criteria for different galaxy types is given in Table \ref{tab:cri}.

\begin{figure*}
\centering
\includegraphics[scale=0.55]{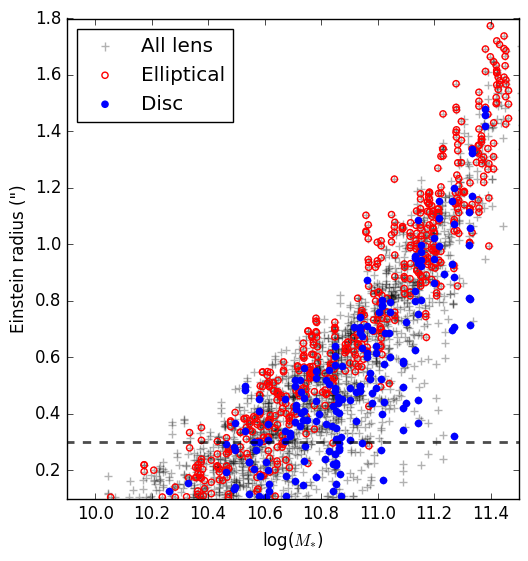}
\includegraphics[scale=0.55]{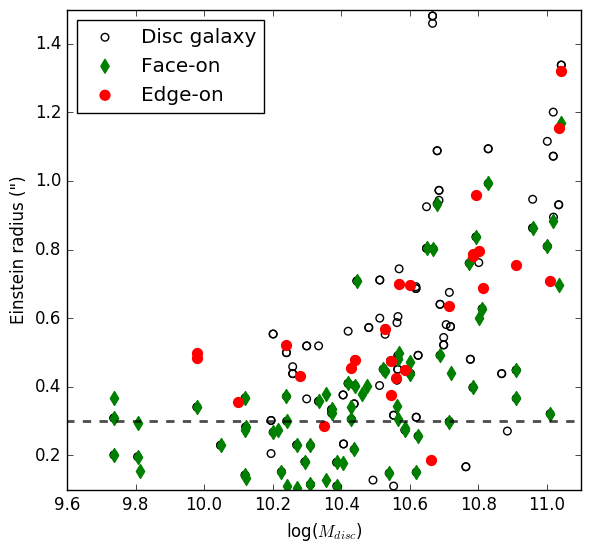}
   \caption{
   {\it Left}: Scatter plot of stellar mass ($h^{-1}~M_{\odot}$) and Einstein radius (arcsecond) for all lenses at redshift $z=0.6$ in Illustris, which are shown as black crosses. All elliptical lenses are shown as red open circles, while disc lenses are shown as blue circles. The Einstein radius cut, $\theta_E > 0.3''$, is shown with the gray dashed line. 
   {\it Right:} Scatter plot of disc mass ($h^{-1}~M_{\odot}$) versus Einstein radius (arcsecond) of all disc lenses as open black circles, face-on disc lenses as green diamonds, and edge-on disc lenses as red circles.
   }\label{fig:sm_sca}
\end{figure*}

The left panel of Figure \ref{fig:sm_sca} shows the distribution of stellar mass versus Einstein radius for disc lenses and elliptical lenses that satisfy the selection criteria as listed in Table \ref{tab:cri}, as well as all other lenses which do not fit to our definition of either disc or elliptical galaxies. The right panel of Figure \ref{fig:sm_sca} shows the distribution of the disc mass versus the Einstein radius for all disc lens candidates in Illustris at $z = 0.6$, where edge-on and face-on discs are labeled as red dots and green diamonds, respectively.  At fixed stellar-disc mass, the edge-on disc lenses tend to have larger Einstein radii than the face-on disc lenses, which indicates that the edge-on disc galaxies are more efficient lenses.
Our analysis focuses on lenses with Einstein radii larger than 0.3 arcseconds, in order to  match the current observational limitations.
Due to computational limitations, for our ray-tracing experiment we randomly select subsets of galaxies of different types. Our complete ray-tracing samples comprise the following four groups: 50 elliptical galaxies, 25 edge-on disc galaxies, 25 face-on disc galaxies, and a sample of 50 disc lenses that includes members of both the edge-on and face-on samples plus disc galaxies with $50 < i <80$.

\begin{table}
\centering
\caption{Summary of the selection criteria. Galaxies of each type must satisfy all the criteria listed.
}
\begin{tabular}{cc}
\hline
\hline
Disc galaxy & Elliptical galaxy \\
\hline
Exponential disc profile  & De Vaucouleur profile \\
S$\grave{e}$rsic index < 2& S$\grave{e}$rsic index > 2\\
Disc star fraction > 40 \% &  Bulge star fraction > 60 \%\\
Face-on: $i < 50^{\circ}$ & \\
Edge-on: $i > 80^{\circ}$ & \\
\hline
\end{tabular}
\label{tab:cri}
\end{table}

\section{Flux-ratio anomalies in Illustris}


Lenses with small opening angles in the fold and cusp configurations are the most sensitive to the perturbations of the gravitational potential that { are responsible for flux-ratio anomalies.
 The relations $R_{\rm fold}$ and $R_{\rm cusp}$, which can be used to describe the anomaly strengths, are given below \citep{Blanford1986,Mao1992,Schneider1992,Zakharov1995,K03}:
\begin{equation}
R_{\rm fold} = \frac{\mu_A + \mu_B}{|\mu_A| + |\mu_B|}~,
\end{equation}
and
\begin{equation}
 R_{\rm cusp} = \frac{\mu_B + \mu_A + \mu_c}{|\mu_A|+|\mu_B|+|\mu_C|}~,
\end{equation}
where $\mu_{A,B}$ in $R_{\rm fold}$ are the magnifications of the two lensed images that make up the merging double, and $\mu_{A,B,C}$ in $R_{\rm cusp}$ are the magnifications of the merging triplet images. The signs of the magnifications indicate image parities. 
As the opening angle, namely $\Delta \phi$ in the cusp configuration  (the opening angle between images A and C) and $\phi_1$ in the fold configuration  (the opening angle between images A and B), approaches zero, $R_{\rm fold}$ and $R_{\rm cusp}$ also approach zero when the lensing mass distribution is smooth and the background object is a point source. The anomaly strength of each lens system can therefore be expressed by its $R_{\rm fold}$ and $R_{\rm cusp}$.  

In order to reduce computational time for the ray-tracing experiment, we randomly draw source positions within a fixed distance from the tangential critical curve where the cusp and fold images form. This distance is fixed to a certain fraction of $r_{\rm caus}$, where $r_{\rm caus}\equiv \sqrt{ab}$, and $a$ and $b$ are half the lengths of the long and short axes of the tangential caustic, respectively. After testing different fractions of $r_{\rm caus}$ on the flux anomaly strength distribution, we select 0.25 $r_{\rm caus}$ for the full ray-tracing run on the mock lenses (see appendix for more discussion.)


\begin{figure}
\includegraphics[scale=0.55]{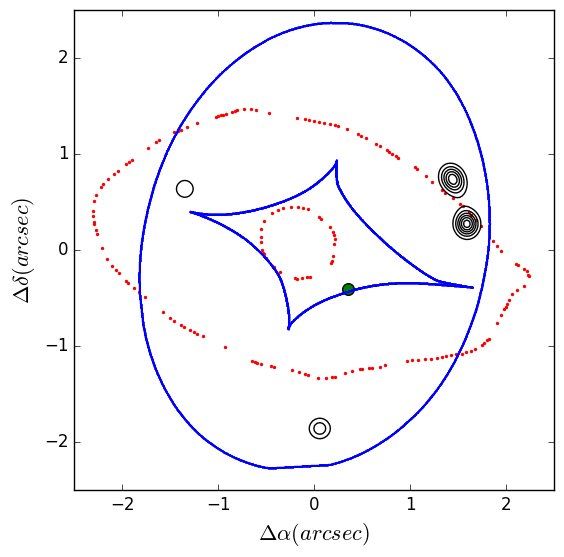}
   \caption{
   An example of a mock lens showing a typical fold configuration from GLAMER with a $\sigma_s = 40$~pc gaussian source and a 50 mas PSF convolution. The lensed images are shown as black contours. The critical curves are shown as red dotted curves and the caustics are shown as blue solid curves. The source position is marked by the green diamond. 
   }\label{fig:raytrace}
\end{figure}

\subsection{Ray-tracing} \label{ssec:raytrace}

We proceed by running a full ray-tracing for the selected simulated galaxies, in order to explore the baryonic effects on flux-ratio anomalies.

The ray-tracing is done with two mock lens sets. From each mock lens, 100 sources are randomly sampled inside the main tangential caustic, within the area chosen so that the distance to the tangential caustic is smaller than $0.25~ r_{\rm caus}$, in order to get fold and cusp image configurations. 
We carry out our calculations in two steps. At the first step we use the ray-tracing code {\sc GLAMER} \citep{Metcalf14} to create a set of idealistic smooth mass models with finite sources. This set of mock lenses provides us with the flux ratio distributions due to a smooth underlying potential with no contribution from either substructure or baryonic components. These mock lenses are generated using the properties of 15 out of the 50 elliptical galaxies in our Illustris ray-tracing samples. In particular, for each selected lens its total mass, velocity dispersion, and axis ratio are used to generate a particle ensemble for a singular isothermal ellipsoid (SIE) halo that has the same mass resolution as the Illustris-1 simulation.

 {\sc GLAMER} calculates the deflection angles, shear and convergence by the tree algorithm \citep{1986Natur.324..446B}, representing each simulation particle with a B-spline in three dimensions as is commonly done in smooth particle hydrodynamics (SPH) simulations.  The size of the particles is set to the distance to the $N_{smooth}$th nearest neighbor where $N_{smooth}$ can be adjusted.  This smoothing scheme provides higher resolution where the particles are dense and it is justified, and lower resolution where the particles are sparse and shot noise would otherwise be a problem.  {\sc GLAMER} also has an adaptive ray-shooting capacity to efficiently resolve small images and facilities for finding caustic curves and representing sources of different kinds.

Since {\sc GLAMER} uses the actual particle distribution for the ray-tracing, without fitting the mass distribution with an analytical profile. This feature allows us to quantify the effect of irregular features in the mass distribution, but it also requires some caution in relation to the particle noise \citep{Xu09, Rau2013}: on the one hand some degree of smoothing is necessary to avoid the particle noise from contributing significantly to the flux-ratio anomalies, on the other hand, one has to be careful as not to smooth out physical small-scale structures. 
We smooth the particle distribution, choosing for each particle a smoothing length that depends on its number of neighbors; after testing the effect of different smoothing lengths we chose $N_{smooth}=64$, which corresponds to $\sim 1 - 3$~kpc near the Einstein radius. Since the characteristic height of a spiral disc is roughly at this physical scale, further increasing the smoothing scale may wash out the edge-on disc structure (see the appendix for more discussion of the particle shot noise.) 
To properly generate the mock quasar lens images, we use a gaussian source with $\sigma_s = 40$~pc, based on typical sizes of narrow-line regions \citep[1--100~pc;][]{Moustakas03}. 

 At the second step we go through the same ray-tracing process on the simulated galaxies that we selected from the Illustris simulation.
We extract the list of particle positions and masses for each of the selected haloes, selecting only the particles belonging to the main halo.
Due to the mass resolution of the Illustris simulation, (self-gravitationally bound) subhalos are only well resolved down to $10^9 M_{\sun}$ ($\sim 500$ dark matter particles), which only covers the relatively high mass end of the subhalo mass function. Indeed as shown in \citet{Xu09,Xu15}, systems with anomalous flux ratios are most sensitive to CDM substructures within a mass range of $10^{7-9}M_{\odot}$ that are projected in the central few kpc region of the main halo\footnote{The upper mass cut is due to the fact that more massive substructures can be ripped apart by tidal interactions and rarely survive in the central region  \citep[see Figure 14 and 15 of][]{Xu09}. The lower mass cut is due to the fact that the perturbation cross section increases with subhalo mass (see Section 5 of \citet{Xu09} and Section 4.5 of \citet{Xu15} for a detailed discussion)}. 
As the Illustris simulation does not provide a subhalo population down to this crucial mass range and since we are only interested in quantifying the effect of baryonic structures, we do not include any simulated subhalos in our ray-tracing analysis.

After the ray-tracing process, the mock lens imaging is convolved with a 50 milli-arcsecond (mas) Gaussian point spread function (PSF), which is typical for the beam size for MERLIN observations obtained at 5~GHz, to obtain realistic flux ratios.
Figure \ref{fig:raytrace} shows an example of a simulated lens  generated by {\sc GLAMER} after the PSF convolution, in which the lensed images are shown with black contours.

\subsection{Flux-ratio anomaly probability distribution} \label{ssec:flux}

Figure \ref{fig:sie} presents the $R_{\rm cusp}$ and $R_{\rm fold}$ distributions that were obtained from ray-tracing through the smooth mass distributions.  The red crosses represent the ray-tracing results, with each cross corresponding to one halo-source pair. The curves in each plot are contours, showing the probabilities of obtaining a given value of \rcusp\ or \rfold\ or larger, as a function of opening angle.  In order to construct the probability contours, the ray-tracing data points are binned every 10 degrees in $\Delta \phi$ and every 5 degrees in $\phi_1$.  These probability distributions are marginally consistent with the smooth results from \citet[see their Figure 4]{Xu15}, where the general smooth lens potentials were modelled as SIEs with observation-motived axis ratios and higher-order multipole perturbations plus random external shear. At small opening angles, our ray-tracing results have slightly higher anomaly strengths compared to \citet{Xu15}, because our sources have finite sizes rather than being point sources.  The observed \rcusp\ and \rfold\ values for real lenses are plotted as black triangles and blue diamonds for elliptical lenses and edge-on disc lenses, respectively (see \citealt{Xu15} table 1 for more observational details).

Figure \ref{fig:cusp} and \ref{fig:fold} show the ray-tracing results for the simulated disc, elliptical, edge-on disc, and face-on disc lenses that include baryons.  We find that the scatter in anomaly strength is larger in the simulated galaxies than in the smooth model lenses for the entire angle range.  The probability of finding low anomaly values ($|R_{\rm cusp}|< 0.3$ and $|R_{\rm fold}|<0.2$) is comparable for the smooth SIE halos and the simulated elliptical lenses.  However, strong anomalies, especially at small opening angles, can only be found when the higher-order moments (mainly from baryonic components) are included.
The strong flux anomalies produced by baryonic components are seen in both elliptical and disc lenses. Furthermore, we see that, among all of our galaxy samples, the edge-on disc lenses produce the widest scatter in anomaly strength. Figure \ref{fig:hist} summaries the ray-tracing results in figure \ref{fig:sie}, \ref{fig:cusp} and \ref{fig:fold} showing the distribution of anomaly values for SIE halos, elliptical lenses, and disk lenses. We find that the strong anomaly events have larger fractions in both elliptical lenses and disk lenses compared to smooth model.

Table \ref{tab:cusp} and \ref{tab:fold} list the probability of finding the flux-ratio anomaly strength for observed lenses. The numbers in parentheses show the probability ranges based on the one sigma uncertainties of the flux-ratio measurements as well as the Poisson noise from the simulated flux-ratio distributions. Note that B0128+437 and B2045+265 are outside of the range of opening angles that we explored for the cusp configuration, while B1608+656 is not listed in the fold configuration for the same reason. Our ray-tracing results show that the baryonic components increase the probability of observing high anomaly strength systems without requiring the presence of dark substructures. Again the systems with edge-on discs are most affected by the baryonic effects.

\begin{figure*}
 \centering															
 \includegraphics[scale=0.55]{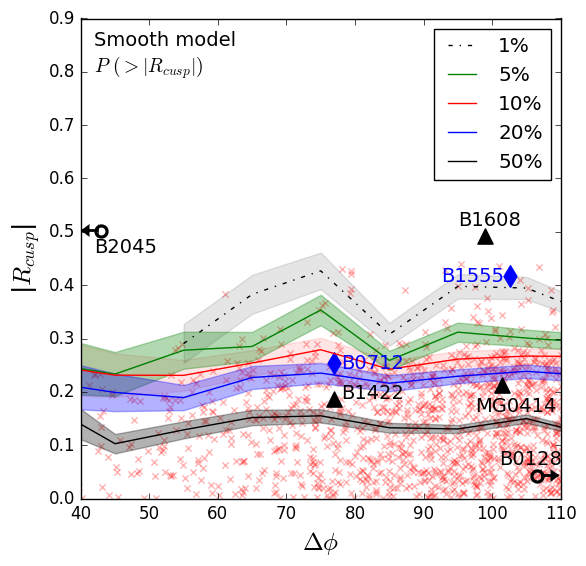}
    \includegraphics[scale=0.55]{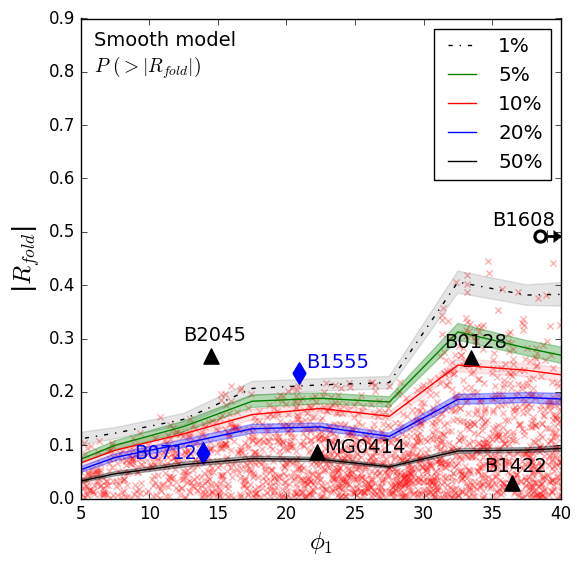}
   \caption{Flux-ratio anomaly strength distributions from the ray-tracing of 10 particle halos that follow the SIE profiles generated by {\sc GLAMER}. The curves represent 1, 5, 10, 20, and 50 per cent probability of finding |$R_{\rm cusp}$| and |$R_{\rm fold}$| larger than a given value for a given opening angle, while the shaded areas represent one sigma uncertainties. The ray-traced data points are shown as red crosses and the observed data points are shown either as blue diamonds (known edge-on discs) or as black triangles. Note that B2045, B0128, and B1608 are outside our complete region of sampling and are represented by open circles. }\label{fig:sie}
\end{figure*}

\begin{figure*}
 \centering
    \includegraphics[scale=0.55]{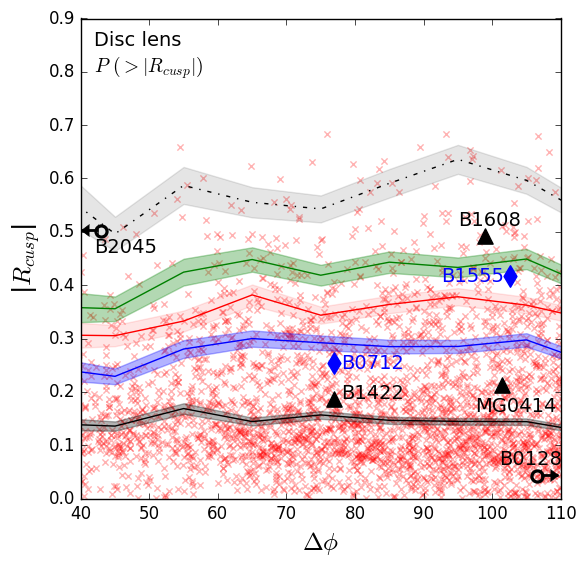}
    \includegraphics[scale=0.55]{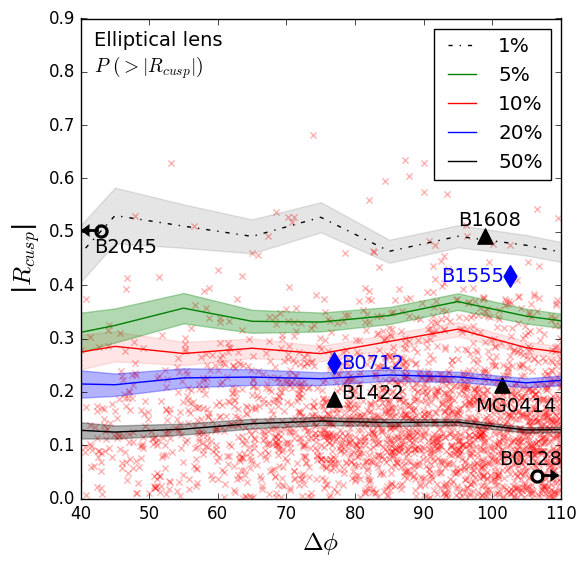}
    \includegraphics[scale=0.55]{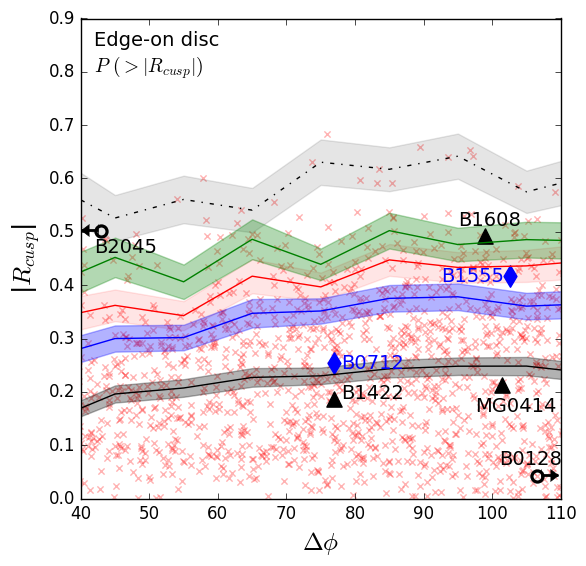}
    \includegraphics[scale=0.55]{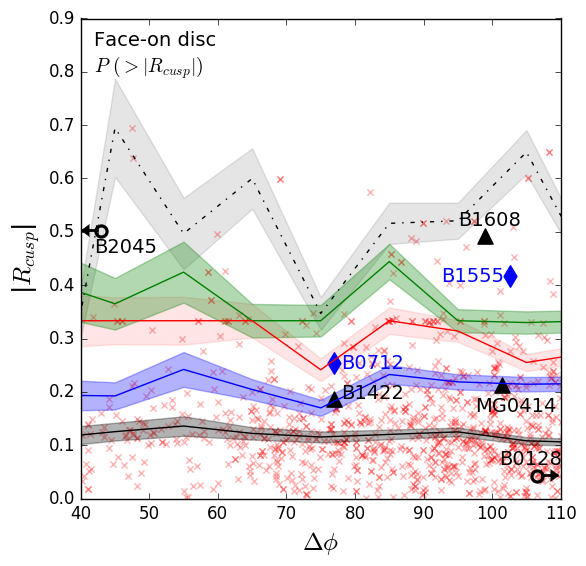}
   \caption{Cusp flux-ratio anomaly strength distribution of four lens galaxy samples (disc, elliptical, edge-on, and face-on) from the ray-tracing results using {\sc GLAMER}. The curves represent 1, 5, 10, 20, and 50 percent probabilities to find |$R_{\rm cusp}$| larger than a given value for a given opening angle ($\Delta \phi (^\circ)$), which the shaded area represents one sigma uncertainty. The ray-traced data points are shown as red crosses and the observed data points are shown as black triangles and the edge-on disc lenses are shown as blue diamonds, respectively. Note that B2045 and B0128 are outside our complete region of sampling and labeled as open circles. }\label{fig:cusp}
\end{figure*}

\begin{figure*}
 \centering
    \includegraphics[scale=0.55]{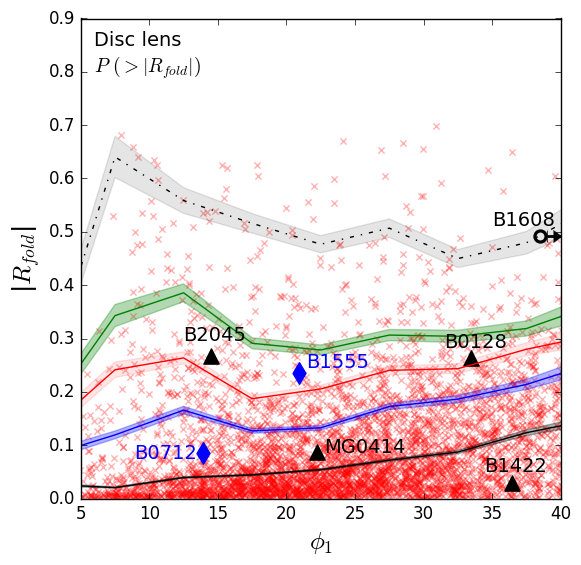}
    \includegraphics[scale=0.55]{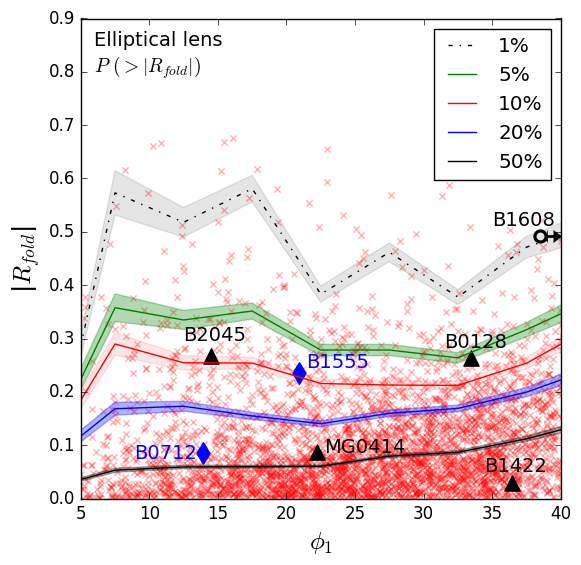}
    \includegraphics[scale=0.55]{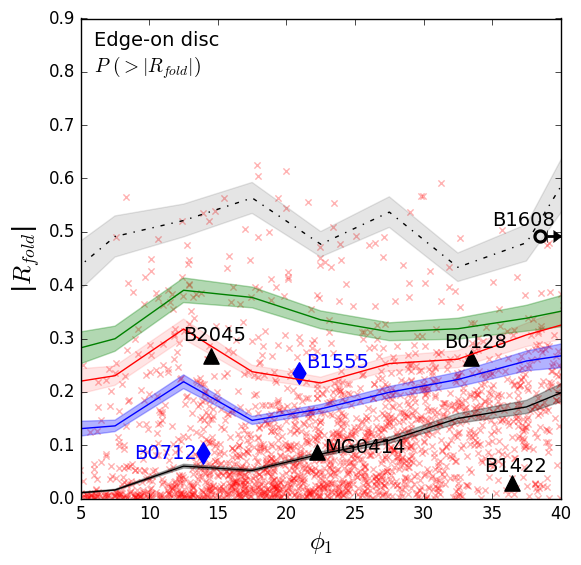}
    \includegraphics[scale=0.55]{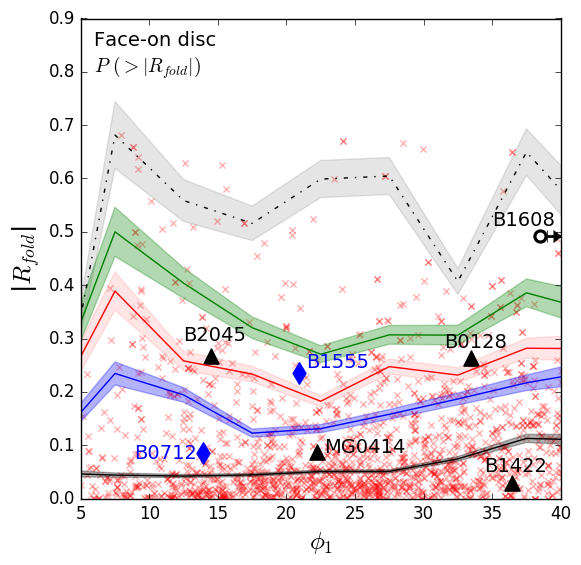}
   \caption{Fold flux-ratio anomaly strength distribution of four lens galaxy samples (disc, elliptical, edge-on, and face-on) from the ray-tracing results using {\sc GLAMER}. The curves represent 1, 5, 10, 20, and 50 percent probabilities to find |$R_{\rm fold}$| larger than a given value for a given opening angle ($ \phi_1 (^\circ)$), which the shaded area represents one sigma uncertainty. The ray-traced data points are shown as red crosses and the observed data points are shown as black triangles and the edge-on disc lenses are shown as blue diamonds, respectively. Note that B1608 is outside our complete region of sampling and labeled as an open circle. }\label{fig:fold}
\end{figure*}

\begin{figure*}
 \centering
    \includegraphics[scale=0.45]{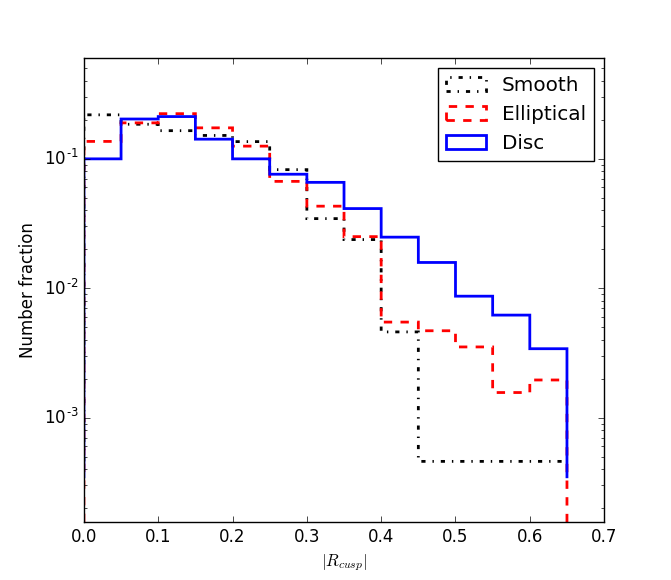}
    \includegraphics[scale=0.45]{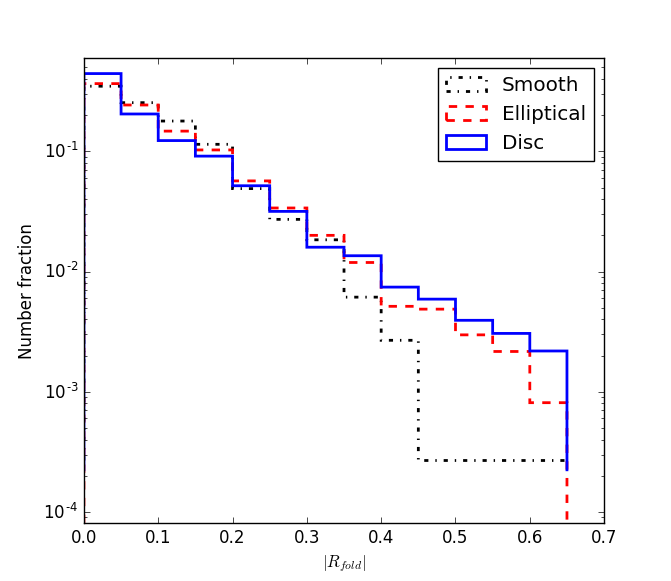}
   \caption{Number fraction histograms showing the distribution of on flux-ratio anomaly strength \rcusp\ ({\it left}) and \rfold ({\it right}) from the ray-tracing results, which black dotted lines representing smooth model, red solid lines represent elliptical lenses, and the blue solid lines representing disk lenses, respectively. Strong anomaly events are more likely to be found with the presence of baryons compared to the smooth model lenses.}\label{fig:hist}
\end{figure*}

\begin{table*}
\centering

\begin{tabular}{cccccccc}
\hline
\hline
Lens &  |$R_{\rm cusp}$|& $\Delta \phi (^\circ)$ &$P_{\rm smooth} (>|R_{\rm cusp}|)$ & $P_{\rm ell} (>|R_{\rm cusp}|)$ & $P_{\rm disc}(>|R_{\rm cusp}|)$ & $P_{\rm edge}(>|R_{\rm cusp}|)$ & $P_{\rm face}(>|R_{\rm cusp}|)$\\
\hline
B1422+231 & 0.187 & 77.0 &0.36 (0.26, 0.46) &0.27 (0.20, 0.34) & 0.37 (0.31, 0.43) & 0.61 (0.54, 0.69) & 0.17 (0.11, 0.22) \\
MG0414+0534 & 0.213 & 101.5 &0.31	(0.10, 0.52) &0.22 (0.11, 0.44) & 0.33 (0.22, 0.49) & 0.57 (0.45, 0.68) & 0.17 (0.08, 0.31)\\
B0712+472 & 0.254 & 76.9 &0.12 (0.08,	0.21) &0.12 (0.06, 0.19) & 0.24 (0.18, 0.29) & 0.43 (0.37, 0.50) & 0.10 (0.03, 0.17)\\
B1555+375& 0.417 & 102.6 &0.01 (0.00, 0.07) &0.02 (0.00, 0.07) & 0.07 (0.05, 0.09) & 0.13 (0.05, 0.20) & 0.02 (0.00, 0.07)\\
B1608+656& 0.492 & 99.0 &0.00 (0.00, 0.08) &0.01 (0.00, 0.06)& 0.03 (0.00, 0.13) & 0.05 (0.01, 0.09) & 0.01 (0.00, 0.04) \\
\hline
\end{tabular}
\caption{Probability of cusp configuration anomaly strength in observed lenses, sorted by |\rcusp|. The numbers in parentheses are the probabilities from the one sigma uncertainty  from flux ratio measurements and Poisson noise of ray-tracing data points. B0128+437 is not listed because its large opening angle ($\Delta \phi = 123^{\circ}$) is out of our sampling region. B2045+265 is similarly not listed due to its small opening angle ($\Delta \phi = 35^{\circ}$). Note that B0712+472 and B1555+375 are confirmed edge-on disc lenses. }
\label{tab:cusp}
\end{table*}

\begin{table*}
\centering

\begin{tabular}{cccccccc}
\hline
\hline
Lens & |$R_{\rm fold}$| & $\phi_1 (^\circ)$ &$P_{\rm smooth} (>|R_{\rm fold}|)$ &$P_{\rm ell} (>|R_{\rm fold}|)$ & $P_{\rm disc}(>|R_{\rm fold}|)$ & $P_{\rm edge}(>|R_{\rm fold}|)$ & $P_{\rm face}(>|R_{\rm fold}|)$\\
\hline
B1422+231 & 0.03 &36.4 &0.82 (0.79, 0.85) &0.85 (0.82, 0.88) & 0.84 (0.82, 0.87) & 0.95 (0.88, 0.95) & 0.82 (0.80, 0.85) \\
B0712+472 & 0.085 &13.9 &0.44 (0.30, 0.60) &0.40 (0.33, 0.50) & 0.31 (0.26, 0.39) & 0.39 (0.32, 0.47) & 0.33 (0.28, 0.41)\\
MG0414+0534 & 0.087 &22.2 &0.45 (0.22, 0.71) &0.35 (0.22, 0.65) & 0.34 (0.20, 0.60) & 0.49 (0.30, 0.71) & 0.31 (0.19, 0.59)\\
B1555+375& 0.235 &20.9 &0.01 (0.00, 0.07) &0.09 (0.05, 0.14) & 0.07 (0.05, 0.12) & 0.10 (0.04, 0.17) & 0.08 (0.05, 0.17)\\
B0128+437 & 0.263 &33.5 &0.11 (0.07, 0.15) & 0.05 (0.01, 0.12) & 0.09 (0.05, 0.16) & 0.12 (0.05, 0.21) & 0.10 (0.05, 0.20) \\
B2045+265& 0.267 &14.5 &0.00 (0.00, 0.06) &0.10 (0.05, 0.18) & 0.08 (0.05, 0.15) & 0.11 (0.05, 0.19) & 0.09 (0.05, 0.20)  \\
\hline
\end{tabular}
\caption{Probability of fold configuration anomaly strength in observed lenses, sorted by |\rfold|. The numbers in parentheses are the probabilities from the one sigma uncertainty  from flux ratio measurements and Poisson noise of ray-tracing data points. B1608+656 is not listed because its large opening angle ($\phi_1 = 48^{\circ}$) is out of our sampling region. Note that B0712+472 and B1555+375 are confirmed edge-on disc lenses.}
\label{tab:fold}
\end{table*}

\section{Discussion}

\subsection{Lens samples from Illustris hydrodynamical simulations}

Interpretation of the simulated galaxies is crucial in studies that try to reproduce observed properties. The Illustris project provides a large sample of galaxies with its large-volume hydrodynamical simulation; together with a few other comparable simulations, such as the EAGLE \citep{schaye15} simulation, it constitutes the state-of-the-art of hydrodynamical simulations, in terms of resolution and complexity of the baryonic physics implementation. \citet{Bottrell2017} compare the galaxies in the Illustris simulation with the Sloan Digital Sky Survey (SDSS) and found that the galaxies in Illustris are larger in size and have a higher fraction of disc-dominated systems than the samples in SDSS with comparable stellar mass. \citet{Vo2014a,Vo2014b} and \citet{Genel2015} also point out that the Illustris simulation only roughly reproduces the stellar mass function and the morphology distributions. We will discuss how these differences limit our sample selection strategy and the interpretation of our analysis results.

In general, the Illustris simulation has stronger feedback processes than other similar simulations - for example the EAGLE simulation \citep{schaye15,Vo2014a,Despali2016}.
The only concern is if the Illustris simulation can provide us a realistic selection of galaxies containing all morphological types. In the Illustris simulation, there are higher fraction of disc systems and also higher fraction of disc-dominated systems among the disc galaxies. Therefore, instead of selecting a galaxy sample with all morphological types, we select only the two extreme categories on the morphological spectrum, massive ellipticals without signs of disc-like structures and late-type disc galaxies with well-developed thin discs. Also, by introducing the kinematic criteria into our selection process, our classification of elliptical and disc galaxies has a better performance than that obtained by using the photometric criteria alone. This is supported by \citet{Bottrell2017}, who showed that the simulated galaxy sample has a better agreement with the observed galaxy sample once the internal kinematics come into the analysis.

In this paper, we focus on the baryon-induced flux-ratio anomalies and compare the strength of the anomalies between elliptical and disc lenses. Although biases in the Illustris simulation may limit our ability to construct a realistic mock survey sample, our data are still adequate for investigating the baryonic effects in the two categories of lens galaxies with the lowest and the strongest perturbation ability.

\subsection{Baryonic effects on flux-ratio anomalies}

The idea that baryonic components such as edge-on discs can generate perturbations to the lensing potential similar to those produced by dark substructures and, therefore, can cause flux-ratio anomalies in lensed quasars has been supported by previous investigations of the B1555+375 and B0712+472 systems \citep{Hsueh2016,Hsueh17}. Additionally, \citet{Gilman2017} have shown that the luminous structure in elliptical lenses may account for 10 to 15 percent of flux-ratio anomalies.
In this work, we select galaxies in the Illustris simulation with well-developed disc structure and compare them to typical elliptical lenses.
By comparing the ray-tracing results of simulated galaxies that include baryons with those from the smooth SIE halos, we find clear signatures of flux-ratio anomalies that have been produced by the higher-order moments that are mainly originated from baryonic structure (see Figure \ref{fig:sie}, \ref{fig:cusp}, and \ref{fig:fold}). 
The scatter in the ray-tracing results (red crosses in Figure \ref{fig:sie}, \ref{fig:cusp}, and \ref{fig:fold}) indicate that the baryons produce more systems that have strong anomaly strengths (i.e., large values of \rcusp\ and \rfold ), and therefore increase the probability of finding such systems. This increase in the strength of the anomalies compared to the smooth model results in Figure \ref{fig:sie} can be clearly seen in the shift of the $5\%$ probability contours, and even, for the disc lenses, of the $20\%$ contours. 

 Events in strong anomaly regions (i.e. |\rcusp| $>$ 0.3 and |\rfold| $>$ 0.2) are particularly rare in the smooth model distributions. By comparing the ray-tracing results  in Figure \ref{fig:sie}, \ref{fig:cusp}, and \ref{fig:fold}, we find that baryons increase the probability of these extreme events by about $8\%$ in elliptical lenses but produce no significant boost in low anomaly regions. For disk lenses, these shifts in the probability curves to higher anomaly values happen on all scales and are most significant at small opening angles. Figure \ref{fig:hist} shows the number fraction distribution of anomaly strengths, which demonstrates the difference between the smooth models and simulated lenses with baryons more clearly. The fraction of elliptical lenses and disk lenses exceed the fraction of SIE halos in strong anomaly regions, and the disk lenses show significantly higher fractions in the most extreme events in |\rcusp|. 
These trends are also observed in Table \ref{tab:cusp} and Table \ref{tab:fold}, showing the probabilities of finding observed lenses among particle halos generated from smooth mass models and mock lenses of different morphological types from the simulation.
It is also interesting to notice that the behavior of baryon-induced anomalies are different from that of substructure-induced anomalies. The former are seen for cusp/fold images with a large range of opening angles, while the latter are only significant at very small opening angles \citep[see Figure 4 of][]{Xu15}. This indicates that, in the large opening angle systems, the flux-ratio anomalies are more likely to be  baryon-induced than substructure-induced, which provides a hint of a method for distinguishing the major sources of flux anomaly in radio-loud lenses. 

While current estimates of substructure abundance and substructure detection using flux-ratio anomalous lenses are based on the assumption that the observed anomalies are generated solely by substructures \citep{Dalal2002,Bradac02,Fadely2012,N14}, we find that the baryonic components can also provide a major contribution to flux-ratio anomalies. Due to the mass resolution of the simulation, we cannot include realistic substructures in this work.  However, our results still provide an explanation for the results seen in \citet{Xu15}, in which they found that dark matter substructures alone are not sufficient to reproduce the observed flux anomaly strengths. Among the four systems which have only a few percent probability for their flux ratios to be solely attributed to the presence of cold dark matter substructures, including baryons makes the observed strong flux anomalies in B0712+472, B1422+231, and B1555+375 more likely; while B2045+265 still has a low probability even in the presence of baryonic components  \citep[However, its high anomaly may be explained by a luminous satellite, see][for more discussion.]{mckean07}  

\subsection{The effect of disc inclination angle}

 In recent investigations of two confirmed edge-on disc lenses, B0712+472 and B1555+375, \citet{Hsueh2016,Hsueh17} show that the inclusion of exponential disc components in the lens model is sufficient to reproduce the observed flux-ratio anomalies without any need to invoke dark matter substructure.  Those results have motivated the work presented in this paper.  
We have found that our general disc lens sample, which includes a range of inclination angles, has a higher probability of causing high flux anomaly systems than the elliptical lens sample.
However, we expect that the disc inclination angle can be an important parameter when considering the effects of baryons in lens systems.
This is borne out by the ray tracing.  As seen in
Figures \ref{fig:cusp} and \ref{fig:fold}, the edge-on discs show larger scatter in flux anomaly strength than both the face-on discs and ellipticals, while the face-on disc lenses produce results similar to what is seen in the elliptical lenses.  These results indicate that the spiral arms in the face-on disc may not be able to cause strong anomalous flux ratios.
Among the morphological types considered here, the edge-on disc lenses produce the strongest flux-ratio anomalies.  In Tables \ref{tab:cusp} and \ref{tab:fold}, the last four columns indicate that baryons increase the probability of finding systems such as B0712+472 and B1555+375, with the edge-on disc lens sample showing the largest increase of the probability of producing large anomalies. 
For example, the probabilities of obtaining the observed flux ratio anomalies in B1555+375 and B0712+472 with smooth mass distributions are only 1\% and 10\%, respectively.  By including edge-on discs, these probabilities rise to 8--15\% and 30--40\%, respectively.
In general, the baryonic components increase the probability of finding large flux-ratio anomalies by $\sim 10 - 20 \%$ for the disc lenses compared to the early-type lenses.  

 In a few cases of face-on lenses, we found that their spiral arms can also distort the critical curves and cause strong flux-ratio anomalies (see Figure \ref{fig:gallery} and Appendix for more discussion).
It is also notable that while all the criteria we used to identify morphological types are designed to select typical examples of each type, the real lens systems can exhibit more complex structures. For example, galaxies with a large thick disc are not selected as ``disc'' galaxies in this work, but their thick discs can still be a more efficient perturber than the stellar structure in the elliptical galaxies. 

In conclusion, our ray-tracing results show that the edge-on disc lenses are the strongest baryonic perturber and have the ability to induce high flux-ratio anomaly in strongly lensed quasars. Their high flux-ratio anomalies can be reproduced by including an exponential disc into the lens model \citep{Hsueh2016,Hsueh17}. These results suggest that the standard ``SIE+shear'' model is not an appropriate analytical model for the edge-on disc lenses. In Figure \ref{fig:sm_hist}, the histograms show that the disc lenses mostly are less massive and have small Einstein radii or, conversely, the smaller the Einstein radius is, the higher the chance that the lensing galaxy is a disc.  However, at these small Einstein radii, it could be difficult to confirm the morphology of the lens galaxy, especially if the lensed quasar images are bright.  In the current sample of radio-loud lensed quasars, three out of eight lenses (B0712+472, B1555+375, and B1933+503) are confirmed to be disc galaxies, B0128+437 is a small separation system and shows possible disc morphology in the Keck AO imaging \citep{lagattuta10}, B1608+656 is a galaxy merger with complicated baryonic structures close to the lensed images, and the rest of them show no obvious sign of discs. Therefore, at least half of the current sample probably requires the consideration of baryonic effects. Moreover, both \citet{Gilman2017} and this current work also show that there is evidence of baryon-induced flux anomalies in elliptical lenses.
This raises the possibility of a bias in the analysis of lens flux ratios, introduced by the lack of knowledge about the baryonic structures in the lens galaxies. It is also notable that disc lenses tend to dominate among small angular separation systems \citep[e.g.][]{Turner84}, which is also shown in Figure 2. As the next generation of large scale surveys, such as the Dark Energy Survey \citep{DES15}, Euclid \citep{Cimatti2012} and the Large Synoptic Survey Telescope \citep{LSST} will be more efficient in finding small angular separation systems, we expect the disc system fraction to increase compared to the current sample.

\subsection{Astrometric anomalies in the Illustris galaxies}

The work of \citet{Gilman2017}, in which mock lenses  included the gravitational perturbation from the luminous mass distribution, shows that the baryonic components can generate flux-ratio anomalies but not astrometric anomalies. 
With our ray-tracing results from the simulated galaxies, we are also able to explore the astrometric anomalies generated by the baryonic components. The astrometric anomaly is defined to be the absolute sum of the deviation between the `real' image positions of the mock lenses and their corresponding best-fit positions produced by a SIE plus external shear model. We analyze the elliptical and edge-on disc mock lenses to see if the morphology of the lens galaxy can also affect astrometric anomalies. We find that about 13 per cent of smooth model predictions have astrometric anomalies larger than $3$ mas, which is the comparable to the precision obtained by very long baseline radio interferometric observations. No obvious difference between the elliptical and edge-on disc lenses in their astrometric anomalies is seen. Our results indicate that the idea of identifying baryonic perturbations by the lack of an observed astrometric anomaly may need to be revisited, since we have seen that baryonic components can also be a cause of astrometric anomalies.

\section{Conclusion}

Motivated by the latest studies on edge-on disc lenses and the resulting baryonic effects on flux-ratio anomalies, we present the first statistical investigation of the baryonic contribution to flux-ratio anomalies from a numerical simulation perspective.  Our ray-tracing results support the idea that the presence of  baryonic structures can induce flux-ratio anomalies (see Figure \ref{fig:sie}, \ref{fig:cusp}, and \ref{fig:fold}), which are clearly seen in both disc lenses and elliptical lenses.  Edge-on disc lenses produced the strongest anomalies among the morphological types that we studied. These results offer a possible explanation for why the observed flux anomaly frequency is higher than what is expected from CDM subhalos as predicted by state-of-the-art N-body simulations \citep{Xu09,Xu15}.  While the magnitudes of the shifts in probability due to the presence of baryons may differ depending on which simulation is used to explore the effect, we expect that the qualitative result of having the baryons increase the likelihood of high values of \rcusp\ and \rfold\ is a generic phenomenon. More detailed investigations can be done on predicting baroynic effects in mock surveys when future hydrodynamical simulations are able to construct more realistic galaxy populations.

 As the monitoring of lensed quasars brings the uncertainties on observed flux ratios down to five per cent \citep{K03}, properly modeling the lens galaxy becomes an increasingly important aspect in the use of flux ratios to provide inferences on substructure abundance.  Our results show that the inclusion of baryonic effects in the modeling is a critical step in this process, especially in the case of disc lenses.  The morphology of the lensing galaxy also needs to be considered in terms of the dark matter substructure where, once again, the presence of a disc can have an important effect.   Studies show that interactions between the disc and substructures can destroy the clumps and cause a lower substructure abundance in disc galaxies compared to halos with elliptical galaxies \citep{Errani2017,DO2010,Yurin2015,Zhu2016}. 
Due to the mass resolution in the Illustris simulation, we cannot include realistic sub-galactic substructures in this work. However, future higher resolution simulations can be used to investigate the contribution of both substructures and baryons to flux-ratio anomalies in the future.

\section*{Acknowledgments}
JWH and CDF would like to thank the Max Planck Institute for Astrophysics for the warm hospitality during their visits, and JWH additionally thanks the HITS for kindly hosting her visit.  CDF acknowledges support from the US National Science foundation grant grant AST-1312329.  DDX would like to thank the Klaus Tschira Foundation.

\bibliographystyle{mnras}
\bibliography{reference}

\begin{thebibliography}{}
\makeatletter
\relax
\def\mn@urlcharsother{\let\do\@makeother \do\$\do\&\do\#\do\^\do\_\do\%\do\~}
\def\mn@doi{\begingroup\mn@urlcharsother \@ifnextchar [ {\mn@doi@}
  {\mn@doi@[]}}
\def\mn@doi@[#1]#2{\def\@tempa{#1}\ifx\@tempa\@empty \href
  {http://dx.doi.org/#2} {doi:#2}\else \href {http://dx.doi.org/#2} {#1}\fi
  \endgroup}
\def\mn@eprint#1#2{\mn@eprint@#1:#2::\@nil}
\def\mn@eprint@arXiv#1{\href {http://arxiv.org/abs/#1} {{\tt arXiv:#1}}}
\def\mn@eprint@dblp#1{\href {http://dblp.uni-trier.de/rec/bibtex/#1.xml}
  {dblp:#1}}
\def\mn@eprint@#1:#2:#3:#4\@nil{\def\@tempa {#1}\def\@tempb {#2}\def\@tempc
  {#3}\ifx \@tempc \@empty \let \@tempc \@tempb \let \@tempb \@tempa \fi \ifx
  \@tempb \@empty \def\@tempb {arXiv}\fi \@ifundefined
  {mn@eprint@\@tempb}{\@tempb:\@tempc}{\expandafter \expandafter \csname
  mn@eprint@\@tempb\endcsname \expandafter{\@tempc}}}

\bibitem[\protect\citeauthoryear{{Abadi}, {Navarro}, {Steinmetz}  \&
  {Eke}}{{Abadi} et~al.}{2003}]{Abadi03}
{Abadi} M.~G.,  {Navarro} J.~F.,  {Steinmetz} M.,   {Eke} V.~R.,  2003, \mn@doi
  [\apj] {10.1086/378316}, \href
  {http://adsabs.harvard.edu/abs/2003ApJ...597...21A} {597, 21}

\bibitem[\protect\citeauthoryear{{Barnes} \& {Hut}}{{Barnes} \&
  {Hut}}{1986}]{1986Natur.324..446B}
{Barnes} J.,  {Hut} P.,  1986, \mn@doi [\nat] {10.1038/324446a0}, \href
  {http://adsabs.harvard.edu/abs/1986Natur.324..446B} {324, 446}

\bibitem[\protect\citeauthoryear{{Blandford} \& {Narayan}}{{Blandford} \&
  {Narayan}}{1986}]{Blanford1986}
{Blandford} R.,  {Narayan} R.,  1986, \mn@doi [\apj] {10.1086/164709}, \href
  {http://adsabs.harvard.edu/abs/1986ApJ...310..568B} {310, 568}

\bibitem[\protect\citeauthoryear{{Bottrell}, {Torrey}, {Simard}  \&
  {Ellison}}{{Bottrell} et~al.}{2017}]{Bottrell2017}
{Bottrell} C.,  {Torrey} P.,  {Simard} L.,   {Ellison} S.~L.,  2017, \mn@doi
  [\mnras] {10.1093/mnras/stx276}, \href
  {http://adsabs.harvard.edu/abs/2017MNRAS.467.2879B} {467, 2879}

\bibitem[\protect\citeauthoryear{{Brada{\v c}}, {Schneider}, {Steinmetz},
  {Lombardi}, {King}  \& {Porcas}}{{Brada{\v c}} et~al.}{2002}]{Bradac02}
{Brada{\v c}} M.,  {Schneider} P.,  {Steinmetz} M.,  {Lombardi} M.,  {King}
  L.~J.,   {Porcas} R.,  2002, \mn@doi [\aap] {10.1051/0004-6361:20020559},
  \href {http://adsabs.harvard.edu/abs/2002A%26A...388..373B} {388, 373}

\bibitem[\protect\citeauthoryear{{Browne}, {Wilkinson}, {Patnaik}  \&
  {Wrobel}}{{Browne} et~al.}{1998}]{JVAS2}
{Browne} I.~W.~A.,  {Wilkinson} P.~N.,  {Patnaik} A.~R.,   {Wrobel} J.~M.,
  1998, \mn@doi [\mnras] {10.1046/j.1365-8711.1998.01072.x}, \href
  {http://adsabs.harvard.edu/abs/1998MNRAS.293..257B} {293, 257}

\bibitem[\protect\citeauthoryear{{Browne} et~al.,}{{Browne}
  et~al.}{2003}]{CLASS2}
{Browne} I.~W.~A.,  et~al., 2003, \mn@doi [\mnras]
  {10.1046/j.1365-8711.2003.06257.x}, \href
  {http://adsabs.harvard.edu/abs/2003MNRAS.341...13B} {341, 13}

\bibitem[\protect\citeauthoryear{{Chen}, {Koushiappas}  \& {Zentner}}{{Chen}
  et~al.}{2011}]{Chen11}
{Chen} J.,  {Koushiappas} S.~M.,   {Zentner} A.~R.,  2011, \mn@doi [\apj]
  {10.1088/0004-637X/741/2/117}, \href
  {http://adsabs.harvard.edu/abs/2011ApJ...741..117C} {741, 117}

\bibitem[\protect\citeauthoryear{{Chiba}, {Minezaki}, {Kashikawa}, {Kataza}  \&
  {Inoue}}{{Chiba} et~al.}{2005}]{Chiba2005}
{Chiba} M.,  {Minezaki} T.,  {Kashikawa} N.,  {Kataza} H.,   {Inoue} K.~T.,
  2005, \mn@doi [\apj] {10.1086/430403}, \href
  {http://adsabs.harvard.edu/abs/2005ApJ...627...53C} {627, 53}

\bibitem[\protect\citeauthoryear{{Cimatti} \& {Scaramella}}{{Cimatti} \&
  {Scaramella}}{2012}]{Cimatti2012}
{Cimatti} A.,  {Scaramella} R.,  2012, Memorie della Societa Astronomica
  Italiana Supplementi, \href
  {http://adsabs.harvard.edu/abs/2012MSAIS..19..314C} {19, 314}

\bibitem[\protect\citeauthoryear{{D'Onghia}, {Vogelsberger}, {Faucher-Giguere}
  \& {Hernquist}}{{D'Onghia} et~al.}{2010}]{DO2010}
{D'Onghia} E.,  {Vogelsberger} M.,  {Faucher-Giguere} C.-A.,   {Hernquist} L.,
  2010, \mn@doi [\apj] {10.1088/0004-637X/725/1/353}, \href
  {http://adsabs.harvard.edu/abs/2010ApJ...725..353D} {725, 353}

\bibitem[\protect\citeauthoryear{{Dalal} \& {Kochanek}}{{Dalal} \&
  {Kochanek}}{2002}]{Dalal2002}
{Dalal} N.,  {Kochanek} C.~S.,  2002, \mn@doi [\apj] {10.1086/340303}, \href
  {http://adsabs.harvard.edu/abs/2002ApJ...572...25D} {572, 25}

\bibitem[\protect\citeauthoryear{{Despali} \& {Vegetti}}{{Despali} \&
  {Vegetti}}{2016}]{Despali2016}
{Despali} G.,  {Vegetti} S.,  2016, preprint, \href
  {http://adsabs.harvard.edu/abs/2016arXiv160806938D} {} (\mn@eprint {arXiv}
  {1608.06938})

\bibitem[\protect\citeauthoryear{{Dobler} \& {Keeton}}{{Dobler} \&
  {Keeton}}{2006}]{Dobler2006}
{Dobler} G.,  {Keeton} C.~R.,  2006, \mn@doi [\mnras]
  {10.1111/j.1365-2966.2005.09809.x}, \href
  {http://adsabs.harvard.edu/abs/2006MNRAS.365.1243D} {365, 1243}

\bibitem[\protect\citeauthoryear{{Errani}, {Pe{\~n}arrubia}, {Laporte}  \&
  {G{\'o}mez}}{{Errani} et~al.}{2017}]{Errani2017}
{Errani} R.,  {Pe{\~n}arrubia} J.,  {Laporte} C.~F.~P.,   {G{\'o}mez} F.~A.,
  2017, \mn@doi [\mnras] {10.1093/mnrasl/slw211}, \href
  {http://adsabs.harvard.edu/abs/2017MNRAS.465L..59E} {465, L59}

\bibitem[\protect\citeauthoryear{{Fadely} \& {Keeton}}{{Fadely} \&
  {Keeton}}{2012}]{Fadely2012}
{Fadely} R.,  {Keeton} C.~R.,  2012, \mn@doi [\mnras]
  {10.1111/j.1365-2966.2011.19729.x}, \href
  {http://adsabs.harvard.edu/abs/2012MNRAS.419..936F} {419, 936}

\bibitem[\protect\citeauthoryear{{Genel} et~al.,}{{Genel}
  et~al.}{2014}]{Genel2014}
{Genel} S.,  et~al., 2014, \mn@doi [\mnras] {10.1093/mnras/stu1654}, \href
  {http://adsabs.harvard.edu/abs/2014MNRAS.445..175G} {445, 175}

\bibitem[\protect\citeauthoryear{{Genel}, {Fall}, {Hernquist}, {Vogelsberger},
  {Snyder}, {Rodriguez-Gomez}, {Sijacki}  \& {Springel}}{{Genel}
  et~al.}{2015}]{Genel2015}
{Genel} S.,  {Fall} S.~M.,  {Hernquist} L.,  {Vogelsberger} M.,  {Snyder}
  G.~F.,  {Rodriguez-Gomez} V.,  {Sijacki} D.,   {Springel} V.,  2015, \mn@doi
  [\apjl] {10.1088/2041-8205/804/2/L40}, \href
  {http://adsabs.harvard.edu/abs/2015ApJ...804L..40G} {804, L40}

\bibitem[\protect\citeauthoryear{{Gilman}, {Agnello}, {Treu}, {Keeton}  \&
  {Nierenberg}}{{Gilman} et~al.}{2016}]{Gilman2016}
{Gilman} D.,  {Agnello} A.,  {Treu} T.,  {Keeton} C.~R.,   {Nierenberg} A.~M.,
  2016, preprint, \href {http://adsabs.harvard.edu/abs/2016arXiv161008525G} {}
  (\mn@eprint {arXiv} {1610.08525})

\bibitem[\protect\citeauthoryear{{Hezaveh} et~al.,}{{Hezaveh}
  et~al.}{2016}]{Hezaveh16}
{Hezaveh} Y.~D.,  et~al., 2016, \mn@doi [\apj] {10.3847/0004-637X/823/1/37},
  \href {http://adsabs.harvard.edu/abs/2016ApJ...823...37H} {823, 37}

\bibitem[\protect\citeauthoryear{{Hinshaw} et~al.,}{{Hinshaw}
  et~al.}{2013}]{wmap9}
{Hinshaw} G.,  et~al., 2013, \mn@doi [\apjs] {10.1088/0067-0049/208/2/19},
  \href {http://adsabs.harvard.edu/abs/2013ApJS..208...19H} {208, 19}

\bibitem[\protect\citeauthoryear{{Hsueh}, {Fassnacht}, {Vegetti}, {McKean},
  {Spingola}, {Auger}, {Koopmans}  \& {Lagattuta}}{{Hsueh}
  et~al.}{2016}]{Hsueh2016}
{Hsueh} J.-W.,  {Fassnacht} C.~D.,  {Vegetti} S.,  {McKean} J.~P.,  {Spingola}
  C.,  {Auger} M.~W.,  {Koopmans} L.~V.~E.,   {Lagattuta} D.~J.,  2016, \mn@doi
  [\mnras] {10.1093/mnrasl/slw146}, \href
  {http://adsabs.harvard.edu/abs/2016MNRAS.463L..51H} {463, L51}

\bibitem[\protect\citeauthoryear{{Hsueh} et~al.,}{{Hsueh}
  et~al.}{2017}]{Hsueh17}
{Hsueh} J.-W.,  et~al., 2017, \mn@doi [\mnras] {10.1093/mnras/stx1082}, \href
  {http://adsabs.harvard.edu/abs/2017MNRAS.469.3713H} {469, 3713}

\bibitem[\protect\citeauthoryear{{Keeton}}{{Keeton}}{2001}]{Kee01}
{Keeton} C.~R.,  2001, ArXiv astro-ph/0102341, \href
  {http://adsabs.harvard.edu/abs/2001astro.ph..2341K} {}

\bibitem[\protect\citeauthoryear{{Kochanek} \& {Dalal}}{{Kochanek} \&
  {Dalal}}{2004}]{KD04}
{Kochanek} C.~S.,  {Dalal} N.,  2004, \mn@doi [\apj] {10.1086/421436}, \href
  {http://adsabs.harvard.edu/abs/2004ApJ...610...69K} {610, 69}

\bibitem[\protect\citeauthoryear{{Koopmans}}{{Koopmans}}{2005}]{K05}
{Koopmans} L.~V.~E.,  2005, \mn@doi [\mnras]
  {10.1111/j.1365-2966.2005.09523.x}, \href
  {http://adsabs.harvard.edu/abs/2005MNRAS.363.1136K} {363, 1136}

\bibitem[\protect\citeauthoryear{{Koopmans} \& {de Bruyn}}{{Koopmans} \& {de
  Bruyn}}{2000}]{koopmans00}
{Koopmans} L.~V.~E.,  {de Bruyn} A.~G.,  2000, \aap, \href
  {http://adsabs.harvard.edu/abs/2000A%26A...358..793K} {358, 793}

\bibitem[\protect\citeauthoryear{{Koopmans} et~al.,}{{Koopmans}
  et~al.}{2003}]{K03}
{Koopmans} L.~V.~E.,  et~al., 2003, \mn@doi [\apj] {10.1086/377434}, \href
  {http://adsabs.harvard.edu/abs/2003ApJ...595..712K} {595, 712}

\bibitem[\protect\citeauthoryear{{LSST Dark Energy Science
  Collaboration}}{{LSST Dark Energy Science Collaboration}}{2012}]{LSST}
{LSST Dark Energy Science Collaboration} 2012, preprint, \href
  {http://adsabs.harvard.edu/abs/2012arXiv1211.0310L} {} (\mn@eprint {arXiv}
  {1211.0310})

\bibitem[\protect\citeauthoryear{{Lagattuta}, {Auger}  \&
  {Fassnacht}}{{Lagattuta} et~al.}{2010}]{lagattuta10}
{Lagattuta} D.~J.,  {Auger} M.~W.,   {Fassnacht} C.~D.,  2010, \mn@doi [\apjl]
  {10.1088/2041-8205/716/2/L185}, \href
  {http://adsabs.harvard.edu/abs/2010ApJ...716L.185L} {716, L185}

\bibitem[\protect\citeauthoryear{{Macci{\`o}} \& {Miranda}}{{Macci{\`o}} \&
  {Miranda}}{2006}]{Maccio06}
{Macci{\`o}} A.~V.,  {Miranda} M.,  2006, \mn@doi [\mnras]
  {10.1111/j.1365-2966.2006.10154.x}, \href
  {http://adsabs.harvard.edu/abs/2006MNRAS.368..599M} {368, 599}

\bibitem[\protect\citeauthoryear{{Mao}}{{Mao}}{1992}]{Mao1992}
{Mao} S.,  1992, \mn@doi [\apjl] {10.1086/186344}, \href
  {http://adsabs.harvard.edu/abs/1992ApJ...389L..41M} {389, L41}

\bibitem[\protect\citeauthoryear{{Mao} \& {Schneider}}{{Mao} \&
  {Schneider}}{1998}]{Mao1998}
{Mao} S.,  {Schneider} P.,  1998, \mn@doi [\mnras]
  {10.1046/j.1365-8711.1998.01319.x}, \href
  {http://adsabs.harvard.edu/abs/1998MNRAS.295..587M} {295, 587}

\bibitem[\protect\citeauthoryear{{Mao}, {Jing}, {Ostriker}  \& {Weller}}{{Mao}
  et~al.}{2004}]{Mao04}
{Mao} S.,  {Jing} Y.,  {Ostriker} J.~P.,   {Weller} J.,  2004, \mn@doi [\apjl]
  {10.1086/383413}, \href {http://adsabs.harvard.edu/abs/2004ApJ...604L...5M}
  {604, L5}

\bibitem[\protect\citeauthoryear{{McCully}, {Keeton}, {Wong}  \&
  {Zabludoff}}{{McCully} et~al.}{2014}]{McCully14}
{McCully} C.,  {Keeton} C.~R.,  {Wong} K.~C.,   {Zabludoff} A.~I.,  2014,
  \mn@doi [\mnras] {10.1093/mnras/stu1316}, \href
  {http://adsabs.harvard.edu/abs/2014MNRAS.443.3631M} {443, 3631}

\bibitem[\protect\citeauthoryear{{McKean} et~al.,}{{McKean}
  et~al.}{2007}]{mckean07}
{McKean} J.~P.,  et~al., 2007, \mn@doi [\mnras]
  {10.1111/j.1365-2966.2007.11744.x}, \href
  {http://adsabs.harvard.edu/abs/2007MNRAS.378..109M} {378, 109}

\bibitem[\protect\citeauthoryear{{Metcalf}}{{Metcalf}}{2005}]{Metcalf05}
{Metcalf} R.~B.,  2005, \mn@doi [\apj] {10.1086/431574}, \href
  {http://adsabs.harvard.edu/abs/2005ApJ...629..673M} {629, 673}

\bibitem[\protect\citeauthoryear{{Metcalf} \& {Amara}}{{Metcalf} \&
  {Amara}}{2012}]{2012MNRAS.419.3414M}
{Metcalf} R.~B.,  {Amara} A.,  2012, \mn@doi [\mnras]
  {10.1111/j.1365-2966.2011.19982.x}, \href
  {http://adsabs.harvard.edu/abs/2012MNRAS.419.3414M} {419, 3414}

\bibitem[\protect\citeauthoryear{{Metcalf} \& {Madau}}{{Metcalf} \&
  {Madau}}{2001}]{metcalf01}
{Metcalf} R.~B.,  {Madau} P.,  2001, \mn@doi [\apj] {10.1086/323695}, \href
  {http://adsabs.harvard.edu/abs/2001ApJ...563....9M} {563, 9}

\bibitem[\protect\citeauthoryear{{Metcalf} \& {Petkova}}{{Metcalf} \&
  {Petkova}}{2014}]{Metcalf14}
{Metcalf} R.~B.,  {Petkova} M.,  2014, \mn@doi [\mnras]
  {10.1093/mnras/stu1859}, \href
  {http://adsabs.harvard.edu/abs/2014MNRAS.445.1942M} {445, 1942}

\bibitem[\protect\citeauthoryear{{Metcalf} \& {Zhao}}{{Metcalf} \&
  {Zhao}}{2002}]{2002ApJ...567L...5M}
{Metcalf} R.~B.,  {Zhao} H.,  2002, \mn@doi [\apjl] {10.1086/339798}, \href
  {http://adsabs.harvard.edu/abs/2002ApJ...567L...5M} {567, L5}

\bibitem[\protect\citeauthoryear{{Minezaki}, {Chiba}, {Kashikawa}, {Inoue}  \&
  {Kataza}}{{Minezaki} et~al.}{2009}]{Minezaki2009}
{Minezaki} T.,  {Chiba} M.,  {Kashikawa} N.,  {Inoue} K.~T.,   {Kataza} H.,
  2009, \mn@doi [\apj] {10.1088/0004-637X/697/1/610}, \href
  {http://adsabs.harvard.edu/abs/2009ApJ...697..610M} {697, 610}

\bibitem[\protect\citeauthoryear{{Mittal}, {Porcas}  \& {Wucknitz}}{{Mittal}
  et~al.}{2007}]{Mittal07}
{Mittal} R.,  {Porcas} R.,   {Wucknitz} O.,  2007, \mn@doi [\aap]
  {10.1051/0004-6361:20066127}, \href
  {http://adsabs.harvard.edu/abs/2007A%26A...465..405M} {465, 405}

\bibitem[\protect\citeauthoryear{{M{\"o}ller}, {Hewett}  \&
  {Blain}}{{M{\"o}ller} et~al.}{2003}]{Moller03}
{M{\"o}ller} O.,  {Hewett} P.,   {Blain} A.~W.,  2003, \mn@doi [\mnras]
  {10.1046/j.1365-8711.2003.06758.x}, \href
  {http://adsabs.harvard.edu/abs/2003MNRAS.345....1M} {345, 1}

\bibitem[\protect\citeauthoryear{{Moustakas} \& {Metcalf}}{{Moustakas} \&
  {Metcalf}}{2003}]{Moustakas03}
{Moustakas} L.~A.,  {Metcalf} R.~B.,  2003, \mn@doi [\mnras]
  {10.1046/j.1365-8711.2003.06055.x}, \href
  {http://adsabs.harvard.edu/abs/2003MNRAS.339..607M} {339, 607}

\bibitem[\protect\citeauthoryear{{Myers} et~al.,}{{Myers}
  et~al.}{2003}]{CLASS1}
{Myers} S.~T.,  et~al., 2003, \mn@doi [\mnras]
  {10.1046/j.1365-8711.2003.06256.x}, \href
  {http://adsabs.harvard.edu/abs/2003MNRAS.341....1M} {341, 1}

\bibitem[\protect\citeauthoryear{{Nelson} et~al.,}{{Nelson}
  et~al.}{2015}]{Nelson2015}
{Nelson} D.,  et~al., 2015, \mn@doi [Astronomy and Computing]
  {10.1016/j.ascom.2015.09.003}, \href
  {http://adsabs.harvard.edu/abs/2015A%26C....13...12N} {13, 12}

\bibitem[\protect\citeauthoryear{{Nierenberg}, {Treu}, {Wright}, {Fassnacht}
  \& {Auger}}{{Nierenberg} et~al.}{2014}]{N14}
{Nierenberg} A.~M.,  {Treu} T.,  {Wright} S.~A.,  {Fassnacht} C.~D.,   {Auger}
  M.~W.,  2014, \mn@doi [\mnras] {10.1093/mnras/stu862}, \href
  {http://adsabs.harvard.edu/abs/2014MNRAS.442.2434N} {442, 2434}

\bibitem[\protect\citeauthoryear{{Nierenberg} et~al.,}{{Nierenberg}
  et~al.}{2017}]{Nierenberg2017}
{Nierenberg} A.~M.,  et~al., 2017, preprint, \href
  {http://adsabs.harvard.edu/abs/2017arXiv170105188N} {} (\mn@eprint {arXiv}
  {1701.05188})

\bibitem[\protect\citeauthoryear{{Patnaik}, {Browne}, {Wilkinson}  \&
  {Wrobel}}{{Patnaik} et~al.}{1992}]{JVAS1}
{Patnaik} A.~R.,  {Browne} I.~W.~A.,  {Wilkinson} P.~N.,   {Wrobel} J.~M.,
  1992, \mn@doi [\mnras] {10.1093/mnras/254.4.655}, \href
  {http://adsabs.harvard.edu/abs/1992MNRAS.254..655P} {254, 655}

\bibitem[\protect\citeauthoryear{{Quadri}, {M{\"o}ller}  \&
  {Natarajan}}{{Quadri} et~al.}{2003}]{Quadri2003}
{Quadri} R.,  {M{\"o}ller} O.,   {Natarajan} P.,  2003, \mn@doi [\apj]
  {10.1086/381216}, \href {http://adsabs.harvard.edu/abs/2003ApJ...597..659Q}
  {597, 659}

\bibitem[\protect\citeauthoryear{{Rau}, {Vegetti}  \& {White}}{{Rau}
  et~al.}{2013}]{Rau2013}
{Rau} S.,  {Vegetti} S.,   {White} S.~D.~M.,  2013, \mn@doi [\mnras]
  {10.1093/mnras/stt043}, \href
  {http://adsabs.harvard.edu/abs/2013MNRAS.430.2232R} {430, 2232}

\bibitem[\protect\citeauthoryear{{Schaye}, {Crain}, {Bower}, {Furlong},
  {Schaller}, {Theuns}, {Dalla Vecchia}  \& {Frenk}}{{Schaye}
  et~al.}{2015}]{schaye15}
{Schaye} J.,  {Crain} R.~A.,  {Bower} R.~G.,  {Furlong} M.,  {Schaller} M.,
  {Theuns} T.,  {Dalla Vecchia} C.,   {Frenk} C.~S. e.~a.,  2015, \mn@doi
  [\mnras] {10.1093/mnras/stu2058}, \href
  {http://adsabs.harvard.edu/abs/2015MNRAS.446..521S} {446, 521}

\bibitem[\protect\citeauthoryear{{Schneider} \& {Weiss}}{{Schneider} \&
  {Weiss}}{1992}]{Schneider1992}
{Schneider} P.,  {Weiss} A.,  1992, \aap, \href
  {http://adsabs.harvard.edu/abs/1992A%26A...260....1S} {260, 1}

\bibitem[\protect\citeauthoryear{{Springel}}{{Springel}}{2010}]{springel10}
{Springel} V.,  2010, \mn@doi [\araa] {10.1146/annurev-astro-081309-130914},
  \href {http://adsabs.harvard.edu/abs/2010ARA%26A..48..391S} {48, 391}

\bibitem[\protect\citeauthoryear{{Teklu}, {Remus}, {Dolag}, {Beck}, {Burkert},
  {Schmidt}, {Schulze}  \& {Steinborn}}{{Teklu} et~al.}{2015}]{Teklu2015}
{Teklu} A.~F.,  {Remus} R.-S.,  {Dolag} K.,  {Beck} A.~M.,  {Burkert} A.,
  {Schmidt} A.~S.,  {Schulze} F.,   {Steinborn} L.~K.,  2015, \mn@doi [\apj]
  {10.1088/0004-637X/812/1/29}, \href
  {http://adsabs.harvard.edu/abs/2015ApJ...812...29T} {812, 29}

\bibitem[\protect\citeauthoryear{{The DES Collaboration} et~al.,}{{The DES
  Collaboration} et~al.}{2015}]{DES15}
{The DES Collaboration} et~al., 2015, preprint, \href
  {http://adsabs.harvard.edu/abs/2015arXiv150302584T} {} (\mn@eprint {arXiv}
  {1503.02584})

\bibitem[\protect\citeauthoryear{{Turner}, {Ostriker}  \& {Gott}}{{Turner}
  et~al.}{1984}]{Turner84}
{Turner} E.~L.,  {Ostriker} J.~P.,   {Gott} III J.~R.,  1984, \mn@doi [\apj]
  {10.1086/162379}, \href {http://adsabs.harvard.edu/abs/1984ApJ...284....1T}
  {284, 1}

\bibitem[\protect\citeauthoryear{{Vegetti} \& {Koopmans}}{{Vegetti} \&
  {Koopmans}}{2009}]{V09}
{Vegetti} S.,  {Koopmans} L.~V.~E.,  2009, \mn@doi [\mnras]
  {10.1111/j.1365-2966.2008.14005.x}, \href
  {http://adsabs.harvard.edu/abs/2009MNRAS.392..945V} {392, 945}

\bibitem[\protect\citeauthoryear{{Vegetti}, {Koopmans}, {Bolton}, {Treu}  \&
  {Gavazzi}}{{Vegetti} et~al.}{2010}]{V10}
{Vegetti} S.,  {Koopmans} L.~V.~E.,  {Bolton} A.,  {Treu} T.,   {Gavazzi} R.,
  2010, \mn@doi [\mnras] {10.1111/j.1365-2966.2010.16865.x}, \href
  {http://adsabs.harvard.edu/abs/2010MNRAS.408.1969V} {408, 1969}

\bibitem[\protect\citeauthoryear{{Vegetti}, {Lagattuta}, {McKean}, {Auger},
  {Fassnacht}  \& {Koopmans}}{{Vegetti} et~al.}{2012}]{V12}
{Vegetti} S.,  {Lagattuta} D.~J.,  {McKean} J.~P.,  {Auger} M.~W.,  {Fassnacht}
  C.~D.,   {Koopmans} L.~V.~E.,  2012, \mn@doi [\nat] {10.1038/nature10669},
  \href {http://adsabs.harvard.edu/abs/2012Natur.481..341V} {481, 341}

\bibitem[\protect\citeauthoryear{{Vegetti}, {Koopmans}, {Auger}, {Treu}  \&
  {Bolton}}{{Vegetti} et~al.}{2014}]{V14a}
{Vegetti} S.,  {Koopmans} L.~V.~E.,  {Auger} M.~W.,  {Treu} T.,   {Bolton}
  A.~S.,  2014, \mn@doi [\mnras] {10.1093/mnras/stu943}, \href
  {http://adsabs.harvard.edu/abs/2014MNRAS.442.2017V} {442, 2017}

\bibitem[\protect\citeauthoryear{{Vogelsberger}, {Genel}, {Sijacki}, {Torrey},
  {Springel}  \& {Hernquist}}{{Vogelsberger} et~al.}{2013}]{Vog2013}
{Vogelsberger} M.,  {Genel} S.,  {Sijacki} D.,  {Torrey} P.,  {Springel} V.,
  {Hernquist} L.,  2013, \mn@doi [\mnras] {10.1093/mnras/stt1789}, \href
  {http://adsabs.harvard.edu/abs/2013MNRAS.436.3031V} {436, 3031}

\bibitem[\protect\citeauthoryear{{Vogelsberger} et~al.,}{{Vogelsberger}
  et~al.}{2014a}]{Vo2014a}
{Vogelsberger} M.,  et~al., 2014a, \mn@doi [\mnras] {10.1093/mnras/stu1536},
  \href {http://adsabs.harvard.edu/abs/2014MNRAS.444.1518V} {444, 1518}

\bibitem[\protect\citeauthoryear{{Vogelsberger} et~al.,}{{Vogelsberger}
  et~al.}{2014b}]{Vo2014b}
{Vogelsberger} M.,  et~al., 2014b, \mn@doi [\nat] {10.1038/nature13316}, \href
  {http://adsabs.harvard.edu/abs/2014Natur.509..177V} {509, 177}

\bibitem[\protect\citeauthoryear{{Wilkinson}, {Browne}, {Patnaik}, {Wrobel}  \&
  {Sorathia}}{{Wilkinson} et~al.}{1998}]{JVAS3}
{Wilkinson} P.~N.,  {Browne} I.~W.~A.,  {Patnaik} A.~R.,  {Wrobel} J.~M.,
  {Sorathia} B.,  1998, \mn@doi [\mnras] {10.1046/j.1365-8711.1998.01941.x},
  \href {http://adsabs.harvard.edu/abs/1998MNRAS.300..790W} {300, 790}

\bibitem[\protect\citeauthoryear{{Winn}, {Rusin}  \& {Kochanek}}{{Winn}
  et~al.}{2004}]{winn04}
{Winn} J.~N.,  {Rusin} D.,   {Kochanek} C.~S.,  2004, \nat, \href
  {http://adsabs.harvard.edu/abs/2004Natur.427..613W} {427, 613}

\bibitem[\protect\citeauthoryear{{Xu} et~al.,}{{Xu} et~al.}{2009}]{Xu09}
{Xu} D.~D.,  et~al., 2009, \mn@doi [\mnras] {10.1111/j.1365-2966.2009.15230.x},
  \href {http://adsabs.harvard.edu/abs/2009MNRAS.398.1235X} {398, 1235}

\bibitem[\protect\citeauthoryear{{Xu}, {Mao}, {Cooper}, {Wang}, {Gao}, {Frenk}
  \& {Springel}}{{Xu} et~al.}{2010}]{Xu10}
{Xu} D.~D.,  {Mao} S.,  {Cooper} A.~P.,  {Wang} J.,  {Gao} L.,  {Frenk} C.~S.,
   {Springel} V.,  2010, \mn@doi [\mnras] {10.1111/j.1365-2966.2010.17235.x},
  \href {http://adsabs.harvard.edu/abs/2010MNRAS.408.1721X} {408, 1721}

\bibitem[\protect\citeauthoryear{{Xu}, {Mao}, {Cooper}, {Gao}, {Frenk},
  {Angulo}  \& {Helly}}{{Xu} et~al.}{2012}]{Xu12}
{Xu} D.~D.,  {Mao} S.,  {Cooper} A.~P.,  {Gao} L.,  {Frenk} C.~S.,  {Angulo}
  R.~E.,   {Helly} J.,  2012, \mn@doi [\mnras]
  {10.1111/j.1365-2966.2012.20484.x}, \href
  {http://adsabs.harvard.edu/abs/2012MNRAS.421.2553X} {421, 2553}

\bibitem[\protect\citeauthoryear{{Xu}, {Sluse}, {Gao}, {Wang}, {Frenk}, {Mao},
  {Schneider}  \& {Springel}}{{Xu} et~al.}{2015}]{Xu15}
{Xu} D.,  {Sluse} D.,  {Gao} L.,  {Wang} J.,  {Frenk} C.,  {Mao} S.,
  {Schneider} P.,   {Springel} V.,  2015, \mn@doi [\mnras]
  {10.1093/mnras/stu2673}, \href
  {http://adsabs.harvard.edu/abs/2015MNRAS.447.3189X} {447, 3189}

\bibitem[\protect\citeauthoryear{{Xu}, {Springel}, {Sluse}, {Schneider},
  {Sonnenfeld}, {Nelson}, {Vogelsberger}  \& {Hernquist}}{{Xu}
  et~al.}{2017}]{Xu17}
{Xu} D.,  {Springel} V.,  {Sluse} D.,  {Schneider} P.,  {Sonnenfeld} A.,
  {Nelson} D.,  {Vogelsberger} M.,   {Hernquist} L.,  2017, \mn@doi [\mnras]
  {10.1093/mnras/stx899}, \href
  {http://adsabs.harvard.edu/abs/2017MNRAS.469.1824X} {469, 1824}

\bibitem[\protect\citeauthoryear{{Yurin} \& {Springel}}{{Yurin} \&
  {Springel}}{2015}]{Yurin2015}
{Yurin} D.,  {Springel} V.,  2015, \mn@doi [\mnras] {10.1093/mnras/stv1454},
  \href {http://adsabs.harvard.edu/abs/2015MNRAS.452.2367Y} {452, 2367}

\bibitem[\protect\citeauthoryear{{Zakharov}}{{Zakharov}}{1995}]{Zakharov1995}
{Zakharov} A.~F.,  1995, \aap, \href
  {http://adsabs.harvard.edu/abs/1995A%26A...293....1Z} {293, 1}

\bibitem[\protect\citeauthoryear{{Zhu}, {Marinacci}, {Maji}, {Li}, {Springel}
  \& {Hernquist}}{{Zhu} et~al.}{2016}]{Zhu2016}
{Zhu} Q.,  {Marinacci} F.,  {Maji} M.,  {Li} Y.,  {Springel} V.,   {Hernquist}
  L.,  2016, \mn@doi [\mnras] {10.1093/mnras/stw374}, \href
  {http://adsabs.harvard.edu/abs/2016MNRAS.458.1559Z} {458, 1559}

\makeatother
\end{thebibliography}

\appendix

\section{Testing the selection area on source plane}

To test if our selected fraction of $r_{\rm caus}$ can reproduce the realistic flux anomaly strength distribution within the opening angle we are interested, we first use the lens modelling code {\sc gravlens} \citep{Kee01} to create analytical distributions where the source positions are generated using different fractions of $r_{\rm caus}$.
 Figure \ref{fig:elp} shows the distributions of $R_{\rm fold}$ vs. $\phi_1$ (top panels) and $R_{\rm cusp}$ vs. $\Delta \phi$ (bottom panels) for an elliptical lens (left panels) and an edge-on disc lens (right panels).  Open circles, solid circles, and crosses represent the distributions for source positions generated within $0.15~r_{\rm caus}$, $0.25~r_{\rm caus}$ and $0.35~r_{\rm caus}$ from the tangential caustic, respectively. As can be seen, the distributions start to diverge beyond certain values of the opening angle, with the divergence occurring at smaller angles for smaller fractions of $r_{\rm caus}$.  For example, for the $0.25 r_{\rm caus}$, the divergence from the $0.35 r_{\rm caus}$ distributions occur at $\phi_1=45^{\circ}$ in the $R_{\rm fold}-\phi_1$ plane, while in the $R_{\rm cusp}-\Delta \phi$ plane the divergence occurs at $\Delta \phi \sim 110^{\circ}$.  
The majority of the observed flux anomaly lenses (see Section \ref{ssec:flux}) have opening angles below these values. 
We therefore set the source region to be within $0.25~r_{\rm caus}$ to the tangential caustic for the full ray-tracing run on the simulated lenses.

\begin{figure*}
\centering
\includegraphics[scale=0.8]{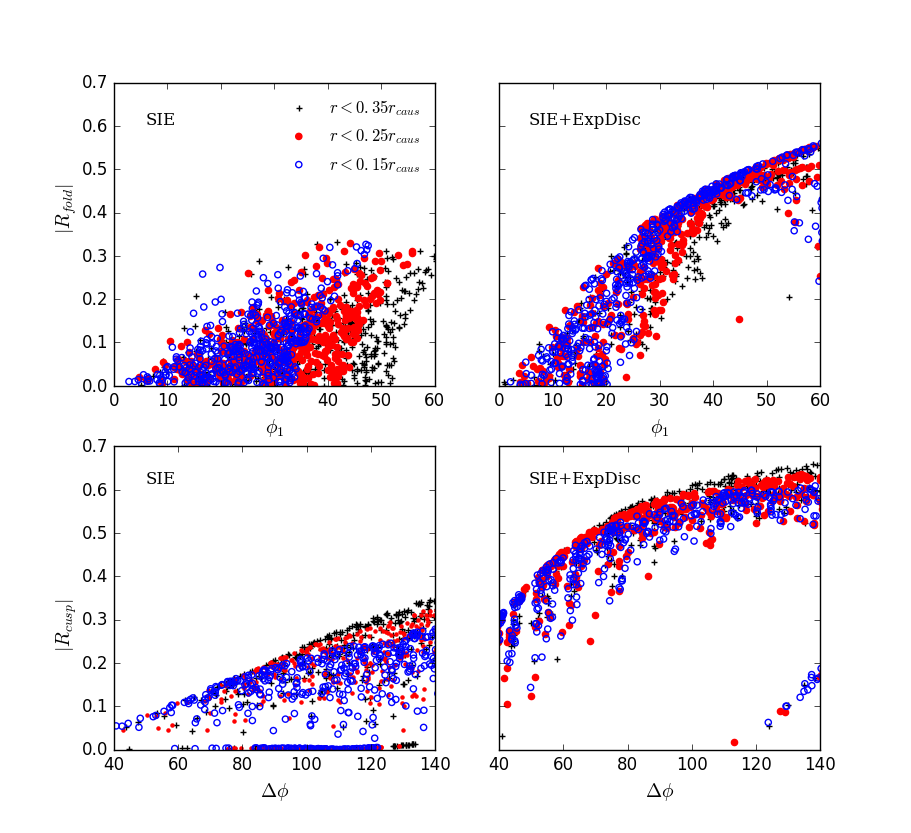}
   \caption{
   Completeness test of flux-ratio anomalies on the analytical model of an elliptical lens and a disc lens, generated by {\sc gravlens}. Left: Singular isothermal ellipsoid model (SIE) with ellipticity $e=0.38$. Right: SIE plus exponential disc (SIE+Expdisc) model with $e=0.76$ and disc mass fraction within Einstein radius equals 15 percent. Each data point represents a point source drawn randomly within the tangential caustic which the distance to the tangential caustic is smaller than $0.15~r_{\rm caus}$ (blue open circles), $0.25~r_{\rm caus}$ (red circles), and $0.35~r_{\rm caus}$ (black crosses). Top: Absolute value of $R_{\rm fold}$ and the fold double opening angle $\Delta \phi$. Bottom: Absolute value of $R_{\rm cusp}$ and the cusp triplet opening angle $\phi_1$.
   }\label{fig:elp}
\end{figure*}

\section{Shot noise test} 
Shot noise is a well known problem that one can encounter when ray-tracing through simulated halos \citep{Rau2013} and it is important to carefully estimate its impact in order to validate our results.
Thus, to understand the contribution from particle noise in our ray-tracing results, we take 10 halos from the Illustris simulation and for each of them, we generate  SIE halos into analytical form and particle ensembles: these last  have the same resolution of the original data, but a smooth particle distribution, so that we can test the effect of shot noise in a setup where there is no irregular structure in the lens halo. We then ray-trace through them with the same setting as we described in \ref{ssec:raytrace}. Both for the analytical and the particle SIE, we use the shape parameters obtained by fitting a SIE to the original data. Figure \ref{fig:shot} shows that the anomalous strength in analytical SIE and particle SIE (which is the smooth model described in \ref{ssec:raytrace}) are at the same level, and so we find that with this smoothing level ($\sim 1 - 3 ~ kpc$ around the Einstein radius), the particle noise does not affect the scattering of flux-ratio anomaly data points nor our statistics. In particular, here both $R_{\rm fold}$  and $R_{\rm cusp}$ are lower than 0.4, while in Figure \ref{fig:cusp} and \ref{fig:fold} higher values are present in all the panels. We therefore conclude that the shot noise does not change our interpretation on the ray-tracing results presented in this paper and that the excess of flux ratio anomalies seen for the simulated elliptical and disk galaxies is not due to particle noise, but to the presence of non-smooth structures in the lens galaxies, such as baryonic components.

\begin{figure*}
 \centering
    \includegraphics[scale=0.35]{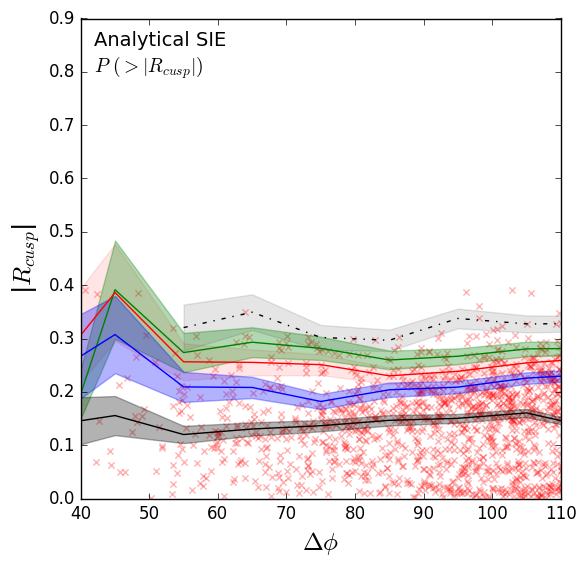}
    \includegraphics[scale=0.35]{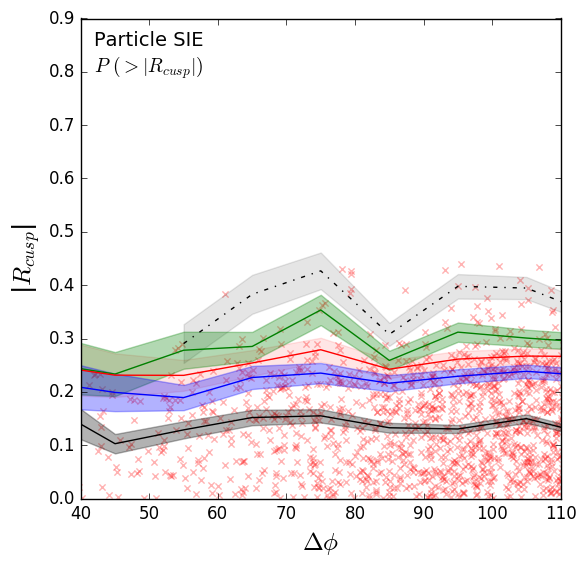}
    \includegraphics[scale=0.35]{glamer_elp_cusp_pd.png}
    \includegraphics[scale=0.35]{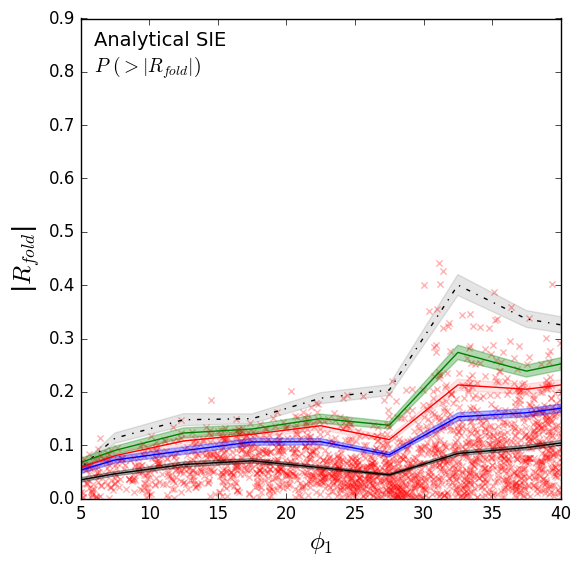}
    \includegraphics[scale=0.35]{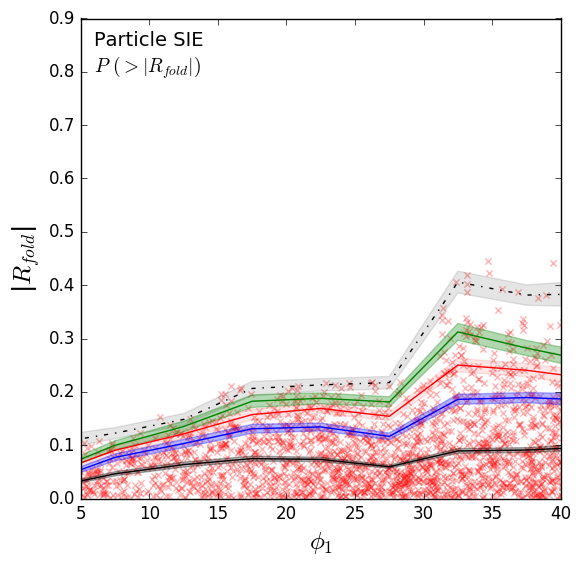}
    \includegraphics[scale=0.35]{glamer_elp_fold_pd.png}
   \caption{
    Flux-ratio anomaly strength distribution of analytical SIE, particle SIE halo (the smooth model results in Fig. \ref{fig:sie}), and the simulated elliptical lens ray-tracing results using {\sc GLAMER}. The curves represent 1, 5, 10, 20, and 50 percent of probability to find |$R_{\rm cusp}$| or |$R_{\rm fold}$|  larger than a given value for a given opening angle ($\Delta \phi$ or $\phi_1 (^\circ)$), which the shaded area represents one sigma uncertainty. The ray-traced data points are shown as red crosses. }\label{fig:shot}
\end{figure*} 

\section{Baryonic effects on the shape of critical curves}

 The baryonic components in the lens galaxy can be the source of perturbations that contribute to the flux-ratio anomalies and also impact the shape of the critical curves. Figure \ref{fig:gallery} demonstrates this effect by showing the young stars (black dots) overlapping with the critical curve (red curve) of a face-on disc lens, an edge-on disc lens, and an elliptical lens. We've seen in a few cases of face-on disc lenses that when the lens galaxy is massive enough, the critical curve can be distorted by the spiral arms. In the case of edge-on disc lenses, the critical curves are mostly elongated shaped due to the stellar disc with a high inclination angle. The elliptical lenses can show some level of distortions in the shape of their critical curves but in general retain their elliptical critical curves.

\begin{figure*}
 \centering
    \includegraphics[scale=0.37]{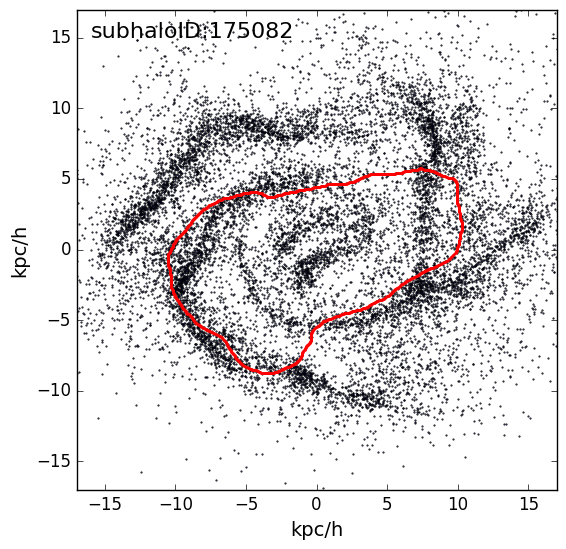}
    \includegraphics[scale=0.37]{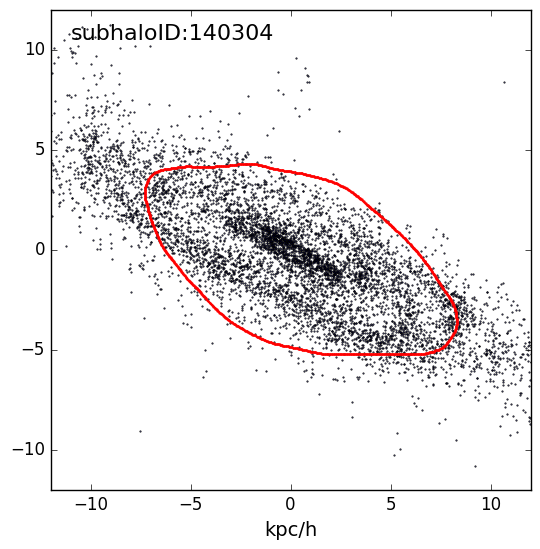}
    \includegraphics[scale=0.37]{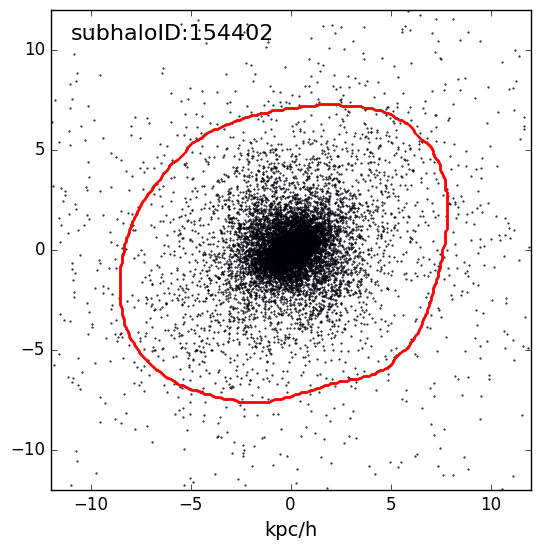}
    \includegraphics[scale=0.27]{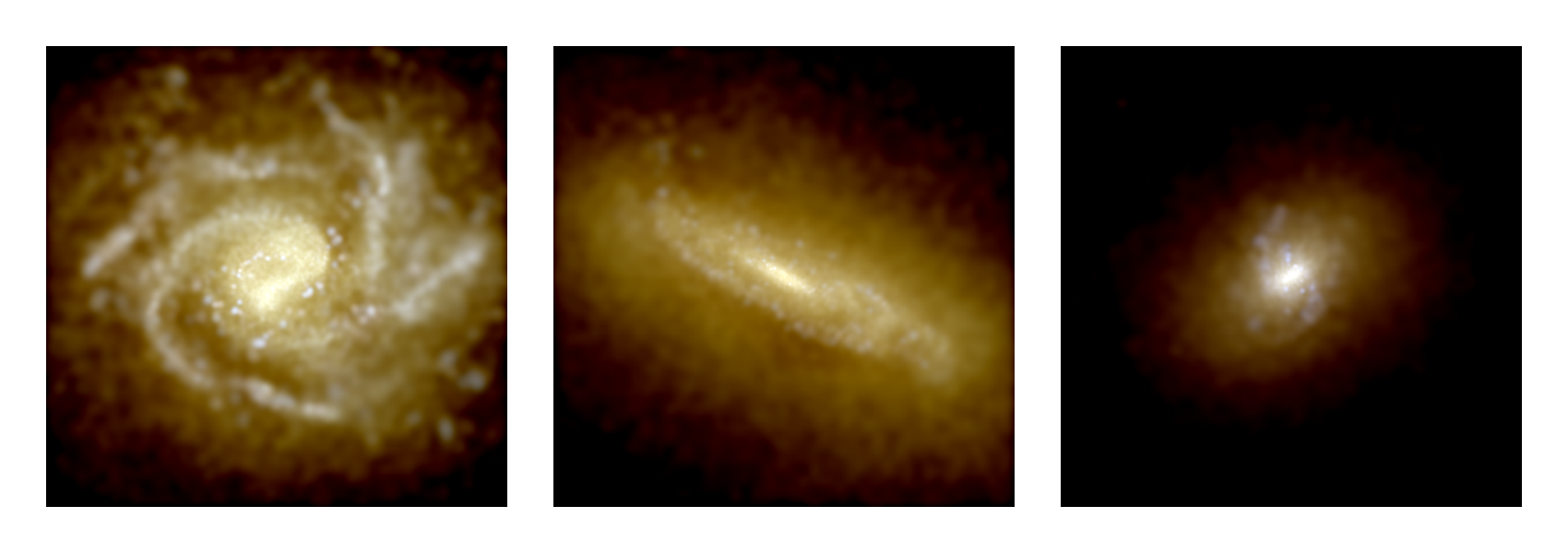}

   \caption{
   {\it Upper}: Young stars (black dots) and critical curves (red curves) of a face-on disc lens, an edge-on disc lens, and an elliptical lens in our ray-tracing samples. The star particles plotted have less than 0.3 Gyr of star formation time in disc galaxies and less than 2 Gyr in the elliptical galaxy. Note that the star particles in the Illustris simulation are not individual stars. The star formation time refers to the time when the star particle was formed once the local gas density excesses the certain threshold. See \citep{Vog2013} for more details. {\it Bottom:} Synthesized images of each lens in the upper row with SDSS g, r, and i filters \citep[see][for more details.]{Xu17} The field of view of images is $15 \times 15~ h^{-1}{\rm kpc}$.
    }\label{fig:gallery}
\end{figure*} 

\label{lastpage}
\end{document}